\documentclass[%
  reprint,
superscriptaddress,
showpacs,preprintnumbers,
nofootinbib,
nobibnotes,
 amsmath,amssymb, 
 aps,
 prl,
 longbibliography,
]{revtex4-1}

\usepackage{cancel}
\usepackage{accents}
\usepackage{mciteplus,slashed}
\usepackage{amssymb,cancel,amsmath}
\usepackage{dcolumn}
\usepackage{bm}
\usepackage[caption=false]{subfig}
\usepackage{appendix}
\usepackage[T1]{fontenc}	
\usepackage{csvsimple}
\usepackage[colorlinks=true,citecolor=blue,linkcolor=blue, allcolors=blue]{hyperref}
\usepackage[section]{placeins}
\usepackage[capitalise]{cleveref}
\usepackage{booktabs}
\usepackage{graphicx}
\usepackage{mathrsfs}
\usepackage[utf8]{inputenc}
\usepackage{siunitx}
\usepackage[dvipsnames]{xcolor}
\usepackage[normalem]{ulem}

\newcommand{\reffig}[1]{Fig.~\ref{#1}}

\newcommand{\reftab}[1]{Table~\ref{#1}}

\def\minerva{MINER$\nu$A\ }

\usepackage{fontawesome} 

\begin{document}

\preprint{\hfill IPPP/18/113/FERMILAB-PUB-18-686-A-ND-PPD-T}

\title{Testing New Physics Explanations of MiniBooNE Anomaly \\at Neutrino Scattering Experiments}

\author{Carlos~A.~Arg\"uelles}
\email{caad@mit.edu} 
\affiliation{Dept.~of Physics, Massachusetts Institute of Technology, Cambridge, MA 02139, USA}
\author{Matheus Hostert}
\email{matheus.hostert@durham.ac.uk}
\affiliation{Institute for Particle Physics Phenomenology, Department of
Physics, Durham University, South Road, Durham DH1 3LE, United Kingdom}
\author{Yu-Dai Tsai}
\email{ytsai@fnal.gov}
\affiliation{Fermilab, Fermi National Accelerator Laboratory, Batavia, IL 60510, USA}

\date{\today}

\begin{abstract}
{\centering{\href{https://github.com/mhostert/DarkNews}{\large\color{BlueViolet}\faGithub}}
\\}
Heavy neutrinos with additional interactions have recently been proposed as an explanation to the MiniBooNE excess. These scenarios often rely on marginally boosted particles to explain the excess angular spectrum, thus predicting large rates at higher-energy neutrino-electron scattering experiments. We place new constraints on this class of models based on neutrino-electron scattering sideband measurements performed at MINER$\nu$A and CHARM-II. A simultaneous explanation of the angular and energy distributions of the MiniBooNE excess in terms of heavy neutrinos with light mediators is severely constrained by our analysis. In general, high-energy neutrino-electron scattering experiments provide strong constraints on explanations of the MiniBooNE observation involving light mediators.
\end{abstract}
 
\maketitle

\textbf{Introduction --} Non-zero neutrino masses have been established in the last twenty years by measurements of neutrino flavor conversion in natural and human-made sources, including long- and short-baseline experiments.
The overwhelming majority of data supports the three-neutrino framework.
Within this framework, we have measured the mixing angles that parametrize the relationship between mass and flavor eigenstates to few-percent-level precision~\cite{Esteban:2018azc}.
The remaining unknowns are the absolute scale of neutrino masses and their origin, the CP-violating phase, and the mass ordering of the neutrinos.
Nevertheless, anomalies in short-baseline accelerator and reactor experiments~\cite{Athanassopoulos:1996jb,Aguilar:2001ty,AguilarArevalo:2007it,Aguilar-Arevalo:2018gpe} challenge this framework and are yet to receive satisfactory explanations.
Minimal extensions of the three-neutrino framework to explain the anomalies introduce the so-called sterile neutrino states, which do not participate in Standard Model (SM) interactions in order to agree with measurements of the Z-boson invisible decay width~\cite{ALEPH:2010aa}.
Unfortunately, these minimal scenarios are disfavoured as they fail to explain all data~\cite{Collin:2016aqd,Capozzi:2016vac,Dentler:2018sju,Diaz:2019fwt}.
This has led the community to explore non-minimal scenarios.
Along this direction, it is interesting to study well-motivated neutrino-mass models that can also explain the short-baseline anomalies and are testable in the laboratory.
In this work, we will investigate a class of neutrino-mass-related models that have been proposed as an explanation of the anomalous observation of $\nu_e$-like events in MiniBooNE~\cite{Aguilar-Arevalo:2018gpe}.

MiniBooNE is a mineral oil Cherenkov detector located in the Booster Neutrino Beam (BNB), at Fermilab~\cite{AguilarArevalo:2008yp,AguilarArevalo:2008qa}.
Using data collected between 2002 and 2017, the experiment has observed an excess of $\nu_e$-like events that is currently in tension with the standard three-neutrino prediction and is beyond statistical doubt at the $4.7 \sigma$ level~\cite{Aguilar-Arevalo:2018gpe}.
While it is possible that the excess is fully or partially due to systematic uncertainties or SM backgrounds~(see, \textit{e.g.},~\cite{AguilarArevalo:2008rc,Aguilar-Arevalo:2012fmn,Hill:2010zy}), many Beyond the Standard Model (BSM) explanations have been put forth.
These new physics (NP) scenarios typically require the existence of new particles, which can:~participate in short-baseline oscillations~\cite{Murayama:2000hm,Strumia:2002fw,Barenboim:2002ah, GonzalezGarcia:2003jq,Barger:2003xm,Sorel:2003hf,Barenboim:2004wu, Zurek:2004vd, Kaplan:2004dq, Pas:2005rb,deGouvea:2006qd,Schwetz:2007cd, Farzan:2008zv,Hollenberg:2009ws,Nelson:2010hz,Akhmedov:2010vy, Diaz:2010ft,Bai:2015ztj, Giunti:2015mwa,Papoulias:2016edm, Moss:2017pur,Carena:2017qhd}, change the neutrino propagation in matter~\cite{Liao:2016reh, Liao:2018mbg,Asaadi:2017bhx,Doring:2018cob}, or be produced in the beam or in the detector and its surroundings~\cite{Gninenko:2009ks,Gninenko:2010pr,Dib:2011jh,McKeen:2010rx,Masip:2012ke, Masip:2011qb,Gninenko:2012rw,Magill:2018jla}.
These models either increase the conversion of muon- to electron-neutrinos or produce electron-neutrino-like signatures in the detector, where in the latter category one typically exploits the fact that the LSND and MiniBooNE are Cherenkov detectors that cannot distinguish between electrons and photons.
Although many MiniBooNE explanations lack a connection to other open problems in particle physics, recent models~\cite{Bertuzzo:2018ftf,Bertuzzo:2018itn,Ballett:2018ynz,Ballett:2019cqp,Ballett:2019pyw} are motivated by neutrino-mass generation via hidden interactions in the heavy-neutrino sector.
In particular, a common prediction of these models is the upscattering of a light neutrino into a heavy neutrino, usually with masses in the tens to hundreds of MeV, which subsequently decays into a pair of electrons.
To reproduce the MiniBooNE excess angular distribution either the heavy neutrino must have moderate boost factors and the pair of electrons produced need to be collimated~\cite{Bertuzzo:2018itn}, or the heavy neutrino two-body decays must be forbidden~\cite{Ballett:2018ynz}.

In this article, we introduce new techniques to probe models that rely on the ambiguity between photons and electrons to explain the MiniBooNE observation, using the dark neutrino model from~\cite{Bertuzzo:2018itn,Bertuzzo:2018ftf} as a benchmark scenario. Our analysis extends to all models with new marginally boosted particles produced in coherent-like neutrino interactions, as they predict large number of events at higher energies~\cite{Gninenko:2009ks,Gninenko:2010pr,Dib:2011jh,McKeen:2010rx,Masip:2012ke,Masip:2011qb,Gninenko:2012rw,Magill:2018jla}.
Thus, our analysis uses high-energy neutrino-electron scattering measurements~\cite{Auerbach:2001wg,Deniz:2009mu,Bellini:2011rx,Park:2013dax,Valencia:2019mkf,Park:2015eqa,Valencia-Rodriguez:2016vkf,DeWinter:1989zg,Geiregat:1992zv,Vilain:1994qy}.
This process is currently used to normalize the neutrino fluxes, due to its well-understood cross section, and has been a fertile ground for light NP searches~\cite{Pospelov:2017kep,Lindner:2018kjo,Magill:2018tbb}.
Here, however, we expand the capability of these measurements to probe BSM-produced photon-like signatures, by developing a new analysis using previously neglected sideband data.
Our technique is complementary to recent searches for coherent single-photon topologies~\cite{Abe:2019cer}.
Since the upscattering process has a threshold of tens to hundreds of MeV, we focus on two high-energy neutrino experiments: \minerva~\cite{Park:2013dax,Valencia:2019mkf,Park:2015eqa,Valencia-Rodriguez:2016vkf}, a scintillator detector in the Neutrinos at the Main Injector (NuMI) beamline at Fermilab, and CHARM-II~\cite{DeWinter:1989zg,Geiregat:1992zv,Vilain:1994qy}, a segmented calorimeter detector at CERN along the Super Proton Synchrotron (SPS) beamline.
These experiments are complementary in the range of neutrino energies they cover and have different background composition.
In all cases a relevant sideband measurement exists, allowing us to take advantage of the excellent particle reconstruction capabilities of \minerva and the precise measurements at CHARM-II to constrain NP.

\begin{figure}[t!]
    \centering
    \includegraphics[width=0.49\textwidth]{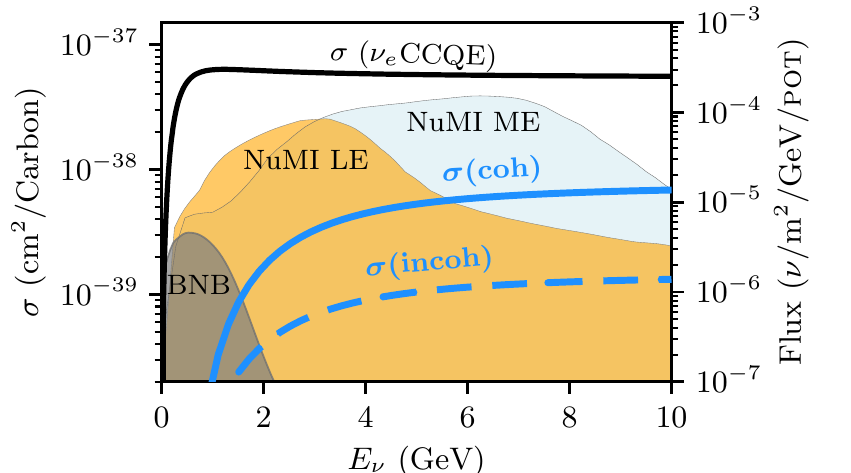}
    \caption{{\textit{Upscattering cross section compared to the quasi-elastic.}} The quasi-elastic cross section on Carbon ($6p^+$) is shown as a function of the neutrino energy (solid black line). The coherent (solid blue) and incoherent (dashed blue) scattering NP cross sections are also shown for the benchmark point of~\cite{Bertuzzo:2018itn}. In the background, we show the BNB flux of $\nu_\mu$ at MiniBooNE (light gray), and the NuMI beam neutrino flux at MINER$\nu$A for the LE (light golden) and ME (light blue) runs in neutrino mode.\label{fig:cross_section}}
\end{figure}
    \textbf{Model --} We consider a minimal realisation of dark neutrino models~\cite{Bertuzzo:2018ftf,Bertuzzo:2018itn,Ballett:2018ynz,Ballett:2019cqp,Ballett:2019pyw} that can explain MiniBooNE.
    This comprises of one Dirac heavy neutrino\footnote{Models with the decay of Majorana particles will lead to greater tension with the angular distribution at MiniBooNE due to their isotropic nature~\cite{Formaggio:1998zn,Balantekin:2018ukw}.}, $\nu_4$, with its associated flavor state, $\nu_D$.
    The dark neutrino $\nu_D$ is charged under a new local U$(1)^\prime$ gauge group, which is part of the particle content and gauge structure needed for mass generation.
    The dark sector is connected to the SM in two ways: kinetic mixing between the new gauge boson and hypercharge, and neutrino mass mixing.
    We start by specifying the kinetic part of the NP Lagrangian
\begin{equation}
\mathscr{L}_{\rm kin} \supset
\;\; \frac{1}{4} \hat{Z}^{\prime}_{\mu \nu} \hat{Z}^{\prime \mu \nu} + \frac{\sin{\chi}}{2} \hat{Z}^{\prime}_{\mu \nu} \hat{B}^{\mu \nu} + \frac{m_{\hat{Z}^\prime}^2}{2} \hat{Z}^{\prime \mu} \hat{Z}^\prime_{\mu},
\end{equation}
where $\hat{Z}^{\prime \mu}$ stands for the new gauge boson field, $\hat{Z}^{\prime \mu\nu}$ its field strength tensor, and $\hat{B}^{\mu \nu}$ the hypercharge field strength tensor.
After usual field redefinitions~\cite{Chun:2010ve}, we arrive at the physical states of the theory.
Working at leading order in $\chi$ and assuming $m_{Z^\prime}^2/m_{Z}^2$ to be small, we can specify the relevant interaction Lagrangian as
\begin{equation}
\mathscr{L}_{\rm int} \supset \;\;g_D \overline{\nu}_D \gamma_\mu \nu_D Z^{\prime \mu}
 + e \varepsilon Z'^{\mu}J^{\rm EM}_{\mu},
\end{equation}
where $J^{\rm EM}_{\mu}$ is the SM electromagnetic current, $g_D$ is the U$(1)^\prime$ gauge coupling assumed to be $\mathcal{O}(1)$, and $\varepsilon \equiv c_{\rm w} \chi$, with $c_{\rm w}$ being the cosine of the weak angle.
Additional terms would be present at higher orders in $\chi$ and mass mixing with the SM $Z$ is also possible, though severely constrained. 
After electroweak symmetry breaking, $\nu_D$ is a superposition of neutrino mass states.
The flavor and mass eigenstates are related via 
\begin{equation}
    \nu_\alpha = \sum^{4}_{i=1} U_{\alpha i}\nu_{i}, \quad (\alpha=e,\mu,\tau,D),
\end{equation}
where $U$ is a $4\times4$ unitary matrix.
It is expected that $|U_{\alpha 4}|$ is small for $\alpha = e, \mu, \tau$, but $|U_{D4}|$ can be of $\mathcal{O}(1)$~\cite{Parke:2015goa,Collin:2016aqd}. 
%
\begin{figure}[t!]
    \centering
    \includegraphics[width=0.50\textwidth]{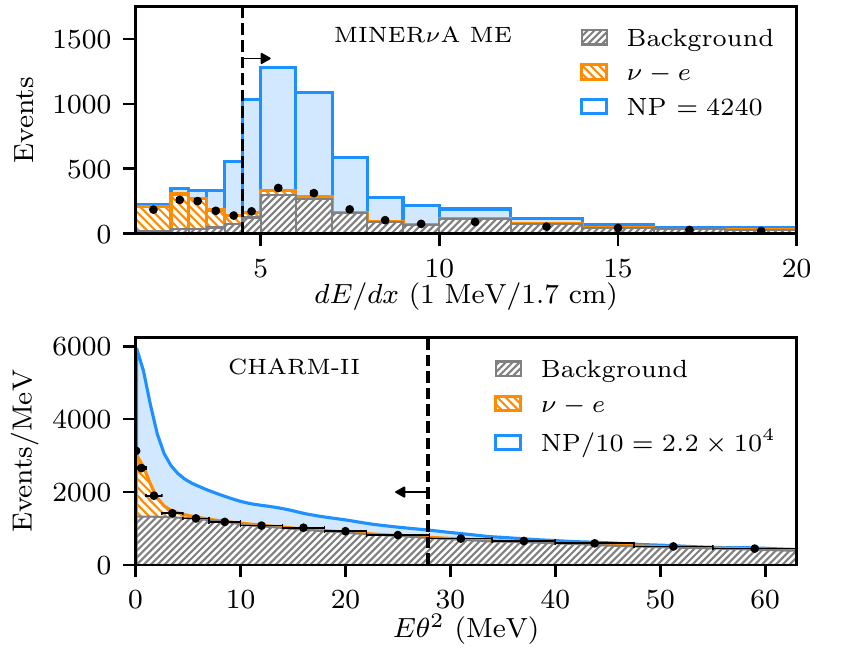}
    \caption{\textit{New physics prediction at \minerva ME and CHARM-II.} Neutrino-electron scattering data in $dE/dx$ at \minerva (top) and in $E\theta^2$ at CHARM-II (bottom). Error bars are too small to be seen. For both experiments, we show the $\nu-e$ signal and total background prediction quoted (after tuning at MINER$\nu$A), as well as the NP prediction (divided by 10 at CHARM-II). The cuts in our analysis our shown as vertical lines. \label{fig:NP_events}}
\end{figure}

\textbf{MiniBooNE signature and region of interest--} The heavy neutrino is produced from an active flavour state upscattering on a nuclear target $A$, $\nu_\alpha A \to \nu_4 A$.
The upscattering cross section is proportional to $\alpha_D \alpha_\textsc{qed}\varepsilon^2 |U_{\alpha 4}|^2$, dominated by $|U_{\mu 4}|$ since all current accelerator neutrino beams are composed mainly of muon neutrinos.
This production can happen off the whole nucleus in a coherent way or off individual nucleons.
For $m_{Z^\prime} \lesssim 100$ MeV, the production will be mainly coherent, but for heavier masses, such as the ones considered in~\cite{Ballett:2018ynz}, incoherent upscattering dominates.
In Fig.~\ref{fig:cross_section}, we show the NP cross section at the benchmark point of~\cite{Bertuzzo:2018itn} and compare it with the quasi-elastic cross section.
By superimposing the cross section on the neutrino fluxes of \minerva and MiniBooNE, we make it explicit that the larger energies at \minerva and CHARM-II are ideal to produce $\nu_4$. Once produced, $\nu_4$ predominantly decays into a neutrino and a dielectron pair, $\nu_4 \to \nu_\alpha e^+ e^-$, either via an on-shell~\cite{Bertuzzo:2018itn} or off-shell~\cite{Ballett:2018ynz} $Z^\prime$ depending on the choice of $m_4$ and $m_{Z^\prime}$.
In this work, we restrict our discussion to the $m_4 > m_{Z^\prime}$ case, where the upscattering is mainly coherent and is followed by a chain of prompt two body decays $\nu_4 \to \nu_\alpha (Z^\prime \to e^+ e^-)$.
The on-shell $Z^\prime$ is required to decay into an overlapping $e^+e^-$ pair, setting a lower bound on its mass of a few MeV. Experimentally, however, $m_{Z^\prime} > 10$ MeV for $e \epsilon \sim 10^{-4}$ to avoid beam dump constraints~\cite{Bauer:2018onh}.
Increasing $m_{Z^\prime}$ increases the ratio of incoherent to coherent events and makes the electron pair less overlapping.
Even though we focus on overlapping $e^+e^-$ pairs, we note that a significant fraction of events would appear as well-separated showers or as a pair of showers with large energy asymmetry, similarly to neutral current (NC) $\pi^0$ events.
The asymmetric events also contribute to the MiniBooNE excess and offer a different target for searches in $\nu-e$ scattering data.

\begin{figure}[t]
    \centering
    \includegraphics[width=3.38in]{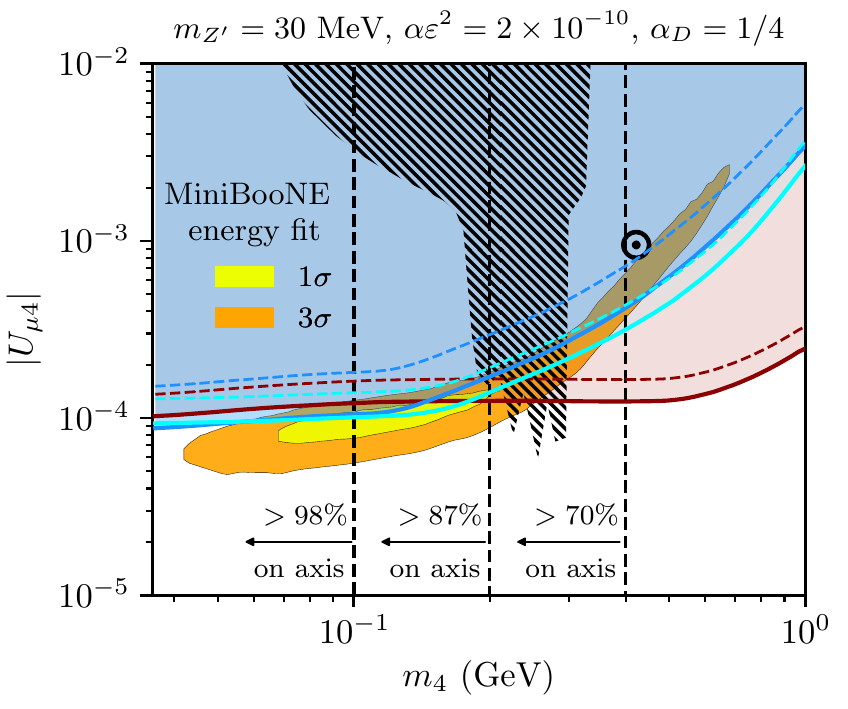}
    \caption{{\textit{New constraints on dark neutrinos as a MiniBooNE explanation.}} The fit to the MiniBooNE energy distribution from~\cite{Bertuzzo:2018itn} is shown as closed yellow (orange) region for one (three) sigma C.L., together with the benchmark point (${\bf\odot}$). Our constraints are shown at 90\% C.L. for \minerva LE in blue (solid -- 30\% background normalization uncertainty, dashed -- conservative 100\% case), for \minerva ME in cyan (solid -- 40\% background normalization uncertainty, dashed -- conservative 100\% case), and for CHARM-II in red (solid -- 3\% background normalization from the sideband constraint, dashed -- conservative 10\% case). Vertical lines show the percentage of excess events at MiniBooNE that lie in the most forward angular bin. Exclusion from heavy neutrino searches is shown as a hatched background. Other relevant assumed parameters are shown above the plot; changing them does not alter our conclusion.\label{fig:final_plot}}
\end{figure}

A fit to the neutrino energy spectrum at MiniBooNE was performed in~\cite{Bertuzzo:2018itn} and is reproduced in~\reffig{fig:final_plot}.
We have performed our own fit to the MiniBooNE energy spectrum using the data release from~\cite{Aguilar-Arevalo:2018gpe}, and our results agree with~\cite{Bertuzzo:2018itn}, when we simulate the signal at MiniBooNE and the analysis cuts in the same way.
This fit leads to preferred values of $m_4$ close to 100 MeV and $|U_{\mu 4 }| \sim 10^{-4}$.
Unfortunately, this energy-only fit neglects the distribution of the excess events as a function of their angle $\theta$ with respect to the beam.
This is important, as the total observed excess contains only $\approx 50\%$ of the events in the most forward bin ($0.8 < \cos{\theta} < 1.0$), with a statistical uncorrelated uncertainty of 5\% on this quantity.  
 
As was recently pointed out in~\cite{Jordan:2018qiy}, few NP scenarios can reproduce the angular distribution of the MiniBooNE excess.
Among these are models where new unstable particles are produced in inelastic collisions in the detector, such as the present case.
Here, large $\theta$ can be achieved by tweaking the mass of the heavy neutrino; the signal becomes less forward as $\nu_4$ becomes heavier.
To show this, we use our dedicated Monte Carlo (MC) simulation to asses the values of $m_4$ preferred by MiniBooNE data~\footnote{Since the released MiniBooNE data do not provide the correlation between angle and energy, and their associated systematics, an energy-angle fit is not possible.}.
For $m_{Z^\prime} = 30$ MeV and $m_4 = 100$, $200$, and $400$ MeV, we find that 98\%, 87\%, and 70\% of the NP events would lie in the most forward bin, respectively. 
We find the predicted angular distribution to be more forward than~\cite{Bertuzzo:2018itn} due to an improved MiniBooNE simulation; see Supplementary Material for details. 
This simulation discrepancy is understood and only strengthens our conclusions.
Thus the relevant region for the MiniBooNE angular distribution is $m_4 \gtrsim 400$ MeV for $m_{Z^\prime} = 30$ MeV.

{\bf Our analysis --} Neutrino-electron scattering measurements predicate their cuts in the following core ideas: no hadronic activity near the interaction vertex, small opening angle from the beam, $E_e \theta^2 \lesssim 2 m_e$, and the requirement that the measured energy deposition, $dE/dx$, be consistent with that of a single electron.
For the NP events, when the coherent process dominates and the mass of the $Z^\prime$ is small, the first two conditions are often satisfied.
However, the requirement of a single-electron-like energy deposition removes a significant fraction of the new-physics induced events. This presents a challenge, as the NP events are mostly overlapping electron pairs and will potentially be removed by the $dE/dx$ cut.
In order to circumvent this problem, we perform our analysis not at the final-cut level, but at an intermediate one.
This is done differently for CHARM-II and MINER$\nu$A: the CHARM-II experiment provides data as a function of $E_e \theta^2$ without the $dE/dx$ cut, and \minerva provides data as a function of the measured $dE/dx$ after analysis cuts on $E_e \theta^2$.

We have developed our own MC simulation for candidate electron pair events in MiniBooNE, MINER$\nu$A and CHARM-II; see the Supplementary Material for more details on detector resolutions, precise signal definition, and resulting distributions.
We only consider the coherent part of the cross section to avoid hadronic-activity cuts, which is conservative.
We also select only events with small energy asymmetries and small opening electron angles.
When required, we assume the mean $dE/dx$ in plastic scintillator to follow the same shape as the NC $\pi^0$ prediction.
Our prediction for new physics events for the BP point is show in Fig.~\ref{fig:NP_events} on top of the \minerva ME and CHARM-II data and MC prediction.


The CHARM-II analysis is mostly based on Fig. 1 of~\cite{Vilain:1994qy}.
This sample is shown as a function of $E\theta^2$ and does not have any cuts on $dE/dx$.
It contains all events with shower energies between $3$ and $24$ GeV, and our final cut on $E\theta^2$ is fixed at $28$ MeV.
For \minerva, the event selection is identical for the low-energy (LE) and medium-energy (ME) analyses~\cite{Park:2015eqa,Valencia:2019mkf}.
The minimum shower energy required is $0.8$ GeV in order to remove the $\pi^0$ background and have reliable angular and energy reconstruction.
Events are kept only when they meet the following angular separation criterion: $E_e \theta^2 < 3.2\times 10^{-3}~{\rm ~GeV ~rad^2}$.
A final cut is applied, ensuring $dE/dx < 4.5~{\rm MeV} / 1.7~{\rm cm}$.
The \minerva analyses use the data outside the previous $dE/dx$ cut to constrain backgrounds. This sideband is defined by all events with $E_e\theta^2 > 5 \times 10^{-3} {\rm ~GeV ~rad^2}$ and $dE/dx < 20~{\rm MeV}/1.7~{\rm cm}$.
Using this sideband measurement, the collaboration tunes their backgrounds by ($0.76$, $0.64$, $1.0$) for ($\nu_e$CCQE, $\nu_\mu$NC, $\nu_\mu$CCQE) processes in the LE mode.
Our LE analysis uses the data shown in Fig. 3 of~\cite{Park:2015eqa} where all the cuts are applied except for the final $dE/dx$ cut.
In our final event selection, we require that the sum of the energy deposited be more than $4.5$ MeV$/ 1.7$ cm, compatible with an $e^+e^-$ pair and yielding an efficiency of $90\%$.

The recent \minerva ME data contains an excess in the region of large $dE/dx$~\cite{Valencia:2019mkf}, where the NP events would lie.
However, this excess is attributed to NC $\pi^0$ events, and grows with the shower energy undershooting the rate require to explain the MiniBooNE anomaly.
With normalization factors as large as 1.7, the collaboration tunes primarily the NC $\pi^0$ prediction in an energy dependent way. 
After tuning, the total NC $\pi^0$ sample corresponds to $20\%$ of the total number of events before the $dE/dx$ cut.

To place our limits, we perform a rate-only analysis by means of a Pearson's $\chi^2$ as test statistic; detailed definition is given in the Supplementary Material.
We incorporate uncertainties in background size and flux normalization as nuisance parameters with Gaussian constraint terms.
For the neutrino-electron scattering and BSM signal, we allow the normalization to scale proportionally to the same flux uncertainty parameter. 
The background term also scales with the flux-uncertainty parameter but has an additional nuisance parameter to account for its unknown size. 
We obtain our constraint as a function of heavy neutrino mass $m_4$, and mixing $|U_{\mu 4}|$ assuming a $\chi^2$ with two degrees of freedom~\cite{Tanabashi:2018oca}.

In our nominal \minerva LE (ME) analysis, we allow for 10\% uncertainty on the flux~\cite{Aliaga:2016oaz}, and 30\% (40\%) uncertainty on the background motivated by the amount of tuning performed on the original backgrounds. Note that the nominal background predictions in the \minerva LE (ME) analysis overpredicts (underpredicts) the data before tuning, and that tuning parameters are measured at the 3\% (5\%) level~\cite{Park:2013dax,Valencia:2019mkf}.
We also perform a background-ignorant analysis in which we assume 100\% uncertainty for the background normalization, which changes our conclusions by only less than a factor of two. This emphasizes the robustness of our \minerva bound, since the NP typically overshoots the low number of events in the sideband. For the benchmark point of~\cite{Bertuzzo:2018itn}, we predict a total signal of 232 (4240) events for \minerva LE (ME).

For CHARM-II, the NP signal lies mostly in a region with small $E\theta^2$.
Thus, we constrain backgrounds using the data from $28 < E\theta^2 < 60$ MeV rad$^2$. This sideband measurement constrains the normalization of the backgrounds in the signal region at the level of $3\%$.
The extrapolation of the shape of the background to the signal region introduces the largest uncertainty in our analysis.
For this reason, we raise the uncertainty of the background normalization from $3\%$ to a conservative $10 \%$ when setting the limits.
Flux uncertainties are assumed to be $4.7\%$ and $5.2\%$ for neutrino and antineutrino mode~\cite{Allaby:1987bb}, respectively, and are applicable to the new-physics signal, $\nu-e$ scattering prediction, and backgrounds. 
Uncertainties in the $\nu-e$ scattering cross sections are expected to be sub-dominant and are neglected in the analysis~\cite{deGouvea:2006hfo}.
For CHARM-II, the NP also yields too many events in the signal region, namely $\approx 2.2\times10^{5}$ events for the benchmark point of~\cite{Bertuzzo:2018itn} in antineutrino mode.
If we lower $|U_{\mu4}| = 10^{-4}$ and $m_4 = 100$ MeV, CHARM-II would still have $\approx 3 \times 10^3$ new physics events. 

{\bf Results and conclusions --} The resulting limits on dark neutrinos using neutrino-electron scattering experiments are shown in the $|U_{\mu 4}|$ vs $m_4$ plane at 90\% confidence level (CL) in~\reffig{fig:final_plot}.
The MiniBooNE fit from~\cite{Bertuzzo:2018itn} is shown, together with vertical lines indicating the percentage of events at MiniBooNE that populate the most forward angular bin.
We have chosen the same values of $\varepsilon$, $\alpha_D$, and $m_{Z^\prime}$ as used in~\cite{Bertuzzo:2018itn}, and shown their benchmark point ($m_4 = 420$ MeV and $|U_{\mu 4}|^2 = 9 \times 10^{-7}$) as a dotted circle.
For these parameters, we can conclude that a good angular distribution at MiniBooNE is in large tension with neutrino-electron scattering data.
We note that the MiniBooNE event rate scales identically to our signal rate in all the couplings, and the dependence on $m_{Z^\prime}$ is subleading due to the typical momentum transfer to the nucleus, provided $m_{Z^\prime} \lesssim 100$ MeV .
This implies that changing the values of these parameters does not modify the overall conclusions of our work.
In addition, for this realization of the model, larger $m_{Z^\prime}$ implies larger values of $m_4$, increasing their impact on neutrino-electron scattering data.
Our \minerva and CHARM-II results are mutually reinforcing given that they impose similar constraints for $m_4 \lesssim 200 $ MeV.
For larger masses, the kinematics of the signal becomes less forward and the production thresholds start being important.
This explains the upturns visible in our bounds, where we observe it first in \minerva and later in CHARM-II as we increase $m_4$, since CHARM-II has higher beam energy.

We emphasize that our analysis is general, and can be adapted to other models. In fact, any MiniBooNE explanation with heavy new particles faces severe constraints from high-energy neutrino-electron scattering data if the signal is free from hadronic activity. This is realised, for instance, in scenarios with heavy neutrinos with dipole interactions~\cite{Gninenko:2009ks,Gninenko:2010pr,Dib:2011jh,McKeen:2010rx,Masip:2012ke,Masip:2011qb,Gninenko:2012rw,Magill:2018jla}. Our bounds can also be adapted to other scenarios with dark neutrinos and heavy mediators~\cite{Ballett:2018ynz,Ballett:2019pyw}. For those, however, we do not expect our bounds to constrain the region of parameter space where the MiniBooNE explanation is viable, since most of the signal at MiniBooNE contains hadronic activity which would be visible at \minerva and CHARM-II.

In the near future, our new analysis strategy could be used in the up-coming \minerva ME results on antineutrino-electron scattering.
The NP cross section, being the same for neutrino and antineutrinos, is thus more prominent on top of backgrounds.
This class of analyses will also greatly benefit from improved calculations and measurements of coherent $\pi^0$ production and single-photon emitting processes.
This is particularly important given the excess seen in the \minerva ME analysis.
A new result can also be obtained by neutrino-electron scattering measurements at NO$\nu$A, which will sample a different kinematic regime as its off-axis beam peaks at lower energies and expects fewer NC~$\pi^0$ events per ton.
Beyond neutrino-electron scattering, the BSM signatures we consider could be lurking in current measurements of $\pi^0$ production, \textit{e.g.}, at MINOS~\cite{Adamson:2016hyz} and MINER$\nu$A~\cite{Wolcott:2016hws}~\footnote{This $\nu_e$CCQE measurement by \minerva observes a significant excess of single photon-like showers attributed to diffractive $\pi^0$ events.
These are abundant in similar realizations of this NP model~\cite{Ballett:2018ynz}.}, and in analyses like the single photon search performed by T2K~\cite{Abe:2019cer}.
 Thus, if dark neutrinos are indeed present in current data, our technique will be crucial to confirm it.

To summarize, a variety of measurements are underway to further lay siege to this explanation of the MiniBooNE observation and, simultaneously, start probing testable neutrino mass generation models, as well as other similar NP signatures.

\section*{Acknowledgements}

We thank Janet Conrad, Kareem Farrag, Alberto Gago, Gordan Krnjaic, Trung Le, Pedro Machado, Kevin Mcfarland, and Jorge Morfin for useful discussions, and Jean DeMerit for carefully proofreading our work.
The authors would like to thank Fermilab for the hospitality at the initial stages of this project.
Also, the authors would like to thank Fermilab Theory Group and the CERN Theory Neutrino Platform for organizing the conference ``Physics Opportunities in the Near DUNE Detector Hall,'' which was essential to the completion of this work.
CAA would especially like to thank Fermilab Center for Neutrino Physics summer visitor program for funding his visit. 
MH's work was supported by Conselho Nacional de Ci\^{e}ncia e Tecnologia (CNPq).
CAA is supported by U.S. National Science Foundation (NSF) grant No. PHY-1801996.
This document was prepared by YDT using the resources of the Fermi National Accelerator Laboratory (Fermilab), a U.S. Department of Energy, Office of Science, HEP User Facility. Fermilab is managed by Fermi Research Alliance, LLC (FRA), acting under Contract No. DE-AC02-07CH11359.

\bibliography{main}

\begin{thebibliography}{86}%
\makeatletter
\providecommand \@ifxundefined [1]{%
 \@ifx{#1\undefined}
}%
\providecommand \@ifnum [1]{%
 \ifnum #1\expandafter \@firstoftwo
 \else \expandafter \@secondoftwo
 \fi
}%
\providecommand \@ifx [1]{%
 \ifx #1\expandafter \@firstoftwo
 \else \expandafter \@secondoftwo
 \fi
}%
\providecommand \natexlab [1]{#1}%
\providecommand \enquote  [1]{``#1''}%
\providecommand \bibnamefont  [1]{#1}%
\providecommand \bibfnamefont [1]{#1}%
\providecommand \citenamefont [1]{#1}%
\providecommand \href@noop [0]{\@secondoftwo}%
\providecommand \href [0]{\begingroup \@sanitize@url \@href}%
\providecommand \@href[1]{\@@startlink{#1}\@@href}%
\providecommand \@@href[1]{\endgroup#1\@@endlink}%
\providecommand \@sanitize@url [0]{\catcode `\\12\catcode `\$12\catcode
  `\&12\catcode `\#12\catcode `\^12\catcode `\_12\catcode `\%12\relax}%
\providecommand \@@startlink[1]{}%
\providecommand \@@endlink[0]{}%
\providecommand \url  [0]{\begingroup\@sanitize@url \@url }%
\providecommand \@url [1]{\endgroup\@href {#1}{\urlprefix }}%
\providecommand \urlprefix  [0]{URL }%
\providecommand \Eprint [0]{\href }%
\providecommand \doibase [0]{http://dx.doi.org/}%
\providecommand \selectlanguage [0]{\@gobble}%
\providecommand \bibinfo  [0]{\@secondoftwo}%
\providecommand \bibfield  [0]{\@secondoftwo}%
\providecommand \translation [1]{[#1]}%
\providecommand \BibitemOpen [0]{}%
\providecommand \bibitemStop [0]{}%
\providecommand \bibitemNoStop [0]{.\EOS\space}%
\providecommand \EOS [0]{\spacefactor3000\relax}%
\providecommand \BibitemShut  [1]{\csname bibitem#1\endcsname}%
\let\auto@bib@innerbib\@empty
\bibitem [{\citenamefont {Esteban}\ \emph {et~al.}(2018)\citenamefont
  {Esteban}, \citenamefont {Gonzalez-Garcia}, \citenamefont
  {Hernandez-Cabezudo}, \citenamefont {Maltoni},\ and\ \citenamefont
  {Schwetz}}]{Esteban:2018azc}%
  \BibitemOpen
  \bibfield  {author} {\bibinfo {author} {\bibfnamefont {Ivan}\ \bibnamefont
  {Esteban}}, \bibinfo {author} {\bibfnamefont {M.~C.}\ \bibnamefont
  {Gonzalez-Garcia}}, \bibinfo {author} {\bibfnamefont {Alvaro}\ \bibnamefont
  {Hernandez-Cabezudo}}, \bibinfo {author} {\bibfnamefont {Michele}\
  \bibnamefont {Maltoni}}, \ and\ \bibinfo {author} {\bibfnamefont {Thomas}\
  \bibnamefont {Schwetz}},\ }\bibfield  {title} {\enquote {\bibinfo {title}
  {{Global analysis of three-flavour neutrino oscillations: synergies and
  tensions in the determination of $\theta_23, \delta_CP$, and the mass
  ordering}},}\ }\href@noop {} {\  (\bibinfo {year} {2018})},\ \Eprint
  {http://arxiv.org/abs/1811.05487} {arXiv:1811.05487 [hep-ph]} \BibitemShut
  {NoStop}%
\bibitem [{\citenamefont {Athanassopoulos}\ \emph {et~al.}(1996)\citenamefont
  {Athanassopoulos} \emph {et~al.}}]{Athanassopoulos:1996jb}%
  \BibitemOpen
  \bibfield  {author} {\bibinfo {author} {\bibfnamefont {C.}~\bibnamefont
  {Athanassopoulos}} \emph {et~al.} (\bibinfo {collaboration} {LSND}),\
  }\bibfield  {title} {\enquote {\bibinfo {title} {{Evidence for
  anti-muon-neutrino ---> anti-electron-neutrino oscillations from the LSND
  experiment at LAMPF}},}\ }\href {\doibase 10.1103/PhysRevLett.77.3082}
  {\bibfield  {journal} {\bibinfo  {journal} {Phys. Rev. Lett.}\ }\textbf
  {\bibinfo {volume} {77}},\ \bibinfo {pages} {3082--3085} (\bibinfo {year}
  {1996})},\ \Eprint {http://arxiv.org/abs/nucl-ex/9605003}
  {arXiv:nucl-ex/9605003 [nucl-ex]} \BibitemShut {NoStop}%
\bibitem [{\citenamefont {Aguilar-Arevalo}\ \emph {et~al.}(2001)\citenamefont
  {Aguilar-Arevalo} \emph {et~al.}}]{Aguilar:2001ty}%
  \BibitemOpen
  \bibfield  {author} {\bibinfo {author} {\bibfnamefont {A.}~\bibnamefont
  {Aguilar-Arevalo}} \emph {et~al.} (\bibinfo {collaboration} {LSND}),\
  }\bibfield  {title} {\enquote {\bibinfo {title} {{Evidence for neutrino
  oscillations from the observation of anti-neutrino(electron) appearance in a
  anti-neutrino(muon) beam}},}\ }\href {\doibase 10.1103/PhysRevD.64.112007}
  {\bibfield  {journal} {\bibinfo  {journal} {Phys. Rev.}\ }\textbf {\bibinfo
  {volume} {D64}},\ \bibinfo {pages} {112007} (\bibinfo {year} {2001})},\
  \Eprint {http://arxiv.org/abs/hep-ex/0104049} {arXiv:hep-ex/0104049 [hep-ex]}
  \BibitemShut {NoStop}%
\bibitem [{\citenamefont {Aguilar-Arevalo}\ \emph {et~al.}(2007)\citenamefont
  {Aguilar-Arevalo} \emph {et~al.}}]{AguilarArevalo:2007it}%
  \BibitemOpen
  \bibfield  {author} {\bibinfo {author} {\bibfnamefont {A.~A.}\ \bibnamefont
  {Aguilar-Arevalo}} \emph {et~al.} (\bibinfo {collaboration} {MiniBooNE}),\
  }\bibfield  {title} {\enquote {\bibinfo {title} {{A Search for electron
  neutrino appearance at the $\Delta m^{2} \sim 1$eV$^{2}$ scale}},}\ }\href
  {\doibase 10.1103/PhysRevLett.98.231801} {\bibfield  {journal} {\bibinfo
  {journal} {Phys. Rev. Lett.}\ }\textbf {\bibinfo {volume} {98}},\ \bibinfo
  {pages} {231801} (\bibinfo {year} {2007})},\ \Eprint
  {http://arxiv.org/abs/0704.1500} {arXiv:0704.1500 [hep-ex]} \BibitemShut
  {NoStop}%
\bibitem [{\citenamefont {Aguilar-Arevalo}\ \emph {et~al.}(2018)\citenamefont
  {Aguilar-Arevalo} \emph {et~al.}}]{Aguilar-Arevalo:2018gpe}%
  \BibitemOpen
  \bibfield  {author} {\bibinfo {author} {\bibfnamefont {A.~A.}\ \bibnamefont
  {Aguilar-Arevalo}} \emph {et~al.} (\bibinfo {collaboration} {MiniBooNE}),\
  }\bibfield  {title} {\enquote {\bibinfo {title} {{Observation of a
  Significant Excess of Electron-Like Events in the MiniBooNE Short-Baseline
  Neutrino Experiment}},}\ }\href@noop {} {\  (\bibinfo {year} {2018})},\
  \Eprint {http://arxiv.org/abs/1805.12028} {arXiv:1805.12028 [hep-ex]}
  \BibitemShut {NoStop}%
\bibitem [{\citenamefont {Group}(2010)}]{ALEPH:2010aa}%
  \BibitemOpen
  \bibfield  {author} {\bibinfo {author} {\bibfnamefont {LEP
  Electroweak~Working}\ \bibnamefont {Group}} (\bibinfo {collaboration} {ALEPH,
  CDF, D0, DELPHI, L3, OPAL, SLD, LEP Electroweak Working Group, Tevatron
  Electroweak Working Group, SLD Electroweak and Heavy Flavour Groups}),\
  }\bibfield  {title} {\enquote {\bibinfo {title} {{Precision Electroweak
  Measurements and Constraints on the Standard Model}},}\ }\href@noop {} {\
  (\bibinfo {year} {2010})},\ \Eprint {http://arxiv.org/abs/1012.2367}
  {arXiv:1012.2367 [hep-ex]} \BibitemShut {NoStop}%
\bibitem [{\citenamefont {{Collin, G. H. and Arg\"uelles, C. A. and Conrad, J.
  M. and Shaevitz, M. H.}}(2016)}]{Collin:2016aqd}%
  \BibitemOpen
  \bibfield  {author} {\bibinfo {author} {\bibnamefont {{Collin, G. H. and
  Arg\"uelles, C. A. and Conrad, J. M. and Shaevitz, M. H.}}},\ }\bibfield
  {title} {\enquote {\bibinfo {title} {{First Constraints on the Complete
  Neutrino Mixing Matrix with a Sterile Neutrino}},}\ }\href@noop {} {\
  (\bibinfo {year} {2016})},\ \Eprint {http://arxiv.org/abs/1607.00011}
  {arXiv:1607.00011 [hep-ph]} \BibitemShut {NoStop}%
\bibitem [{\citenamefont {Capozzi}\ \emph {et~al.}(2017)\citenamefont
  {Capozzi}, \citenamefont {Giunti}, \citenamefont {Laveder},\ and\
  \citenamefont {Palazzo}}]{Capozzi:2016vac}%
  \BibitemOpen
  \bibfield  {author} {\bibinfo {author} {\bibfnamefont {Francesco}\
  \bibnamefont {Capozzi}}, \bibinfo {author} {\bibfnamefont {Carlo}\
  \bibnamefont {Giunti}}, \bibinfo {author} {\bibfnamefont {Marco}\
  \bibnamefont {Laveder}}, \ and\ \bibinfo {author} {\bibfnamefont {Antonio}\
  \bibnamefont {Palazzo}},\ }\bibfield  {title} {\enquote {\bibinfo {title}
  {{Joint short- and long-baseline constraints on light sterile neutrinos}},}\
  }\href {\doibase 10.1103/PhysRevD.95.033006} {\bibfield  {journal} {\bibinfo
  {journal} {Phys. Rev.}\ }\textbf {\bibinfo {volume} {D95}},\ \bibinfo {pages}
  {033006} (\bibinfo {year} {2017})},\ \Eprint
  {http://arxiv.org/abs/1612.07764} {arXiv:1612.07764 [hep-ph]} \BibitemShut
  {NoStop}%
\bibitem [{\citenamefont {Dentler}\ \emph {et~al.}(2018)\citenamefont
  {Dentler}, \citenamefont {Hernández-Cabezudo}, \citenamefont {Kopp},
  \citenamefont {Machado}, \citenamefont {Maltoni}, \citenamefont
  {Martinez-Soler},\ and\ \citenamefont {Schwetz}}]{Dentler:2018sju}%
  \BibitemOpen
  \bibfield  {author} {\bibinfo {author} {\bibfnamefont {Mona}\ \bibnamefont
  {Dentler}}, \bibinfo {author} {\bibfnamefont {Álvaro}\ \bibnamefont
  {Hernández-Cabezudo}}, \bibinfo {author} {\bibfnamefont {Joachim}\
  \bibnamefont {Kopp}}, \bibinfo {author} {\bibfnamefont {Pedro A.~N.}\
  \bibnamefont {Machado}}, \bibinfo {author} {\bibfnamefont {Michele}\
  \bibnamefont {Maltoni}}, \bibinfo {author} {\bibfnamefont {Ivan}\
  \bibnamefont {Martinez-Soler}}, \ and\ \bibinfo {author} {\bibfnamefont
  {Thomas}\ \bibnamefont {Schwetz}},\ }\bibfield  {title} {\enquote {\bibinfo
  {title} {{Updated Global Analysis of Neutrino Oscillations in the Presence of
  eV-Scale Sterile Neutrinos}},}\ }\href {\doibase 10.1007/JHEP08(2018)010}
  {\bibfield  {journal} {\bibinfo  {journal} {JHEP}\ }\textbf {\bibinfo
  {volume} {08}},\ \bibinfo {pages} {010} (\bibinfo {year} {2018})},\ \Eprint
  {http://arxiv.org/abs/1803.10661} {arXiv:1803.10661 [hep-ph]} \BibitemShut
  {NoStop}%
\bibitem [{\citenamefont {Diaz}\ \emph {et~al.}(2019)\citenamefont {Diaz},
  \citenamefont {Argüelles}, \citenamefont {Collin}, \citenamefont {Conrad},\
  and\ \citenamefont {Shaevitz}}]{Diaz:2019fwt}%
  \BibitemOpen
  \bibfield  {author} {\bibinfo {author} {\bibfnamefont {A.}~\bibnamefont
  {Diaz}}, \bibinfo {author} {\bibfnamefont {C.~A.}\ \bibnamefont
  {Argüelles}}, \bibinfo {author} {\bibfnamefont {G.~H.}\ \bibnamefont
  {Collin}}, \bibinfo {author} {\bibfnamefont {J.~M.}\ \bibnamefont {Conrad}},
  \ and\ \bibinfo {author} {\bibfnamefont {M.~H.}\ \bibnamefont {Shaevitz}},\
  }\bibfield  {title} {\enquote {\bibinfo {title} {{Where Are We With Light
  Sterile Neutrinos?}}}\ }\href@noop {} {\  (\bibinfo {year} {2019})},\ \Eprint
  {http://arxiv.org/abs/1906.00045} {arXiv:1906.00045 [hep-ex]} \BibitemShut
  {NoStop}%
\bibitem [{\citenamefont {Aguilar-Arevalo}\ \emph
  {et~al.}(2009{\natexlab{a}})\citenamefont {Aguilar-Arevalo} \emph
  {et~al.}}]{AguilarArevalo:2008yp}%
  \BibitemOpen
  \bibfield  {author} {\bibinfo {author} {\bibfnamefont {A.~A.}\ \bibnamefont
  {Aguilar-Arevalo}} \emph {et~al.} (\bibinfo {collaboration} {MiniBooNE}),\
  }\bibfield  {title} {\enquote {\bibinfo {title} {{The Neutrino Flux
  prediction at MiniBooNE}},}\ }\href {\doibase 10.1103/PhysRevD.79.072002}
  {\bibfield  {journal} {\bibinfo  {journal} {Phys. Rev.}\ }\textbf {\bibinfo
  {volume} {D79}},\ \bibinfo {pages} {072002} (\bibinfo {year}
  {2009}{\natexlab{a}})},\ \Eprint {http://arxiv.org/abs/0806.1449}
  {arXiv:0806.1449 [hep-ex]} \BibitemShut {NoStop}%
\bibitem [{\citenamefont {Aguilar-Arevalo}\ \emph
  {et~al.}(2009{\natexlab{b}})\citenamefont {Aguilar-Arevalo} \emph
  {et~al.}}]{AguilarArevalo:2008qa}%
  \BibitemOpen
  \bibfield  {author} {\bibinfo {author} {\bibfnamefont {A.~A.}\ \bibnamefont
  {Aguilar-Arevalo}} \emph {et~al.} (\bibinfo {collaboration} {MiniBooNE}),\
  }\bibfield  {title} {\enquote {\bibinfo {title} {{The MiniBooNE Detector}},}\
  }\href {\doibase 10.1016/j.nima.2008.10.028} {\bibfield  {journal} {\bibinfo
  {journal} {Nucl. Instrum. Meth.}\ }\textbf {\bibinfo {volume} {A599}},\
  \bibinfo {pages} {28--46} (\bibinfo {year} {2009}{\natexlab{b}})},\ \Eprint
  {http://arxiv.org/abs/0806.4201} {arXiv:0806.4201 [hep-ex]} \BibitemShut
  {NoStop}%
\bibitem [{\citenamefont {Aguilar-Arevalo}\ \emph
  {et~al.}(2009{\natexlab{c}})\citenamefont {Aguilar-Arevalo} \emph
  {et~al.}}]{AguilarArevalo:2008rc}%
  \BibitemOpen
  \bibfield  {author} {\bibinfo {author} {\bibfnamefont {A.~A.}\ \bibnamefont
  {Aguilar-Arevalo}} \emph {et~al.} (\bibinfo {collaboration} {MiniBooNE}),\
  }\bibfield  {title} {\enquote {\bibinfo {title} {{Unexplained Excess of
  Electron-Like Events From a 1-GeV Neutrino Beam}},}\ }\href {\doibase
  10.1103/PhysRevLett.102.101802} {\bibfield  {journal} {\bibinfo  {journal}
  {Phys. Rev. Lett.}\ }\textbf {\bibinfo {volume} {102}},\ \bibinfo {pages}
  {101802} (\bibinfo {year} {2009}{\natexlab{c}})},\ \Eprint
  {http://arxiv.org/abs/0812.2243} {arXiv:0812.2243 [hep-ex]} \BibitemShut
  {NoStop}%
\bibitem [{\citenamefont {Aguilar-Arevalo}\ \emph {et~al.}(2012)\citenamefont
  {Aguilar-Arevalo} \emph {et~al.}}]{Aguilar-Arevalo:2012fmn}%
  \BibitemOpen
  \bibfield  {author} {\bibinfo {author} {\bibfnamefont {A.~A.}\ \bibnamefont
  {Aguilar-Arevalo}} \emph {et~al.} (\bibinfo {collaboration} {MiniBooNE}),\
  }\bibfield  {title} {\enquote {\bibinfo {title} {{A Combined $\nu_\mu
  \rightarrow \nu_e$ and $\bar \nu_\mu \rightarrow \bar \nu_e$ Oscillation
  Analysis of the MiniBooNE Excesses}},}\ \ }(\bibinfo {year} {2012})\ \Eprint
  {http://arxiv.org/abs/1207.4809} {arXiv:1207.4809 [hep-ex]} \BibitemShut
  {NoStop}%
\bibitem [{\citenamefont {Hill}(2011)}]{Hill:2010zy}%
  \BibitemOpen
  \bibfield  {author} {\bibinfo {author} {\bibfnamefont {Richard~J.}\
  \bibnamefont {Hill}},\ }\bibfield  {title} {\enquote {\bibinfo {title} {{On
  the single photon background to $\nu_e$ appearance at MiniBooNE}},}\ }\href
  {\doibase 10.1103/PhysRevD.84.017501} {\bibfield  {journal} {\bibinfo
  {journal} {Phys. Rev.}\ }\textbf {\bibinfo {volume} {D84}},\ \bibinfo {pages}
  {017501} (\bibinfo {year} {2011})},\ \Eprint {http://arxiv.org/abs/1002.4215}
  {arXiv:1002.4215 [hep-ph]} \BibitemShut {NoStop}%
\bibitem [{\citenamefont {Murayama}\ and\ \citenamefont
  {Yanagida}(2001)}]{Murayama:2000hm}%
  \BibitemOpen
  \bibfield  {author} {\bibinfo {author} {\bibfnamefont {Hitoshi}\ \bibnamefont
  {Murayama}}\ and\ \bibinfo {author} {\bibfnamefont {T.}~\bibnamefont
  {Yanagida}},\ }\bibfield  {title} {\enquote {\bibinfo {title} {{LSND,
  SN1987A, and CPT violation}},}\ }\href {\doibase
  10.1016/S0370-2693(01)01136-4} {\bibfield  {journal} {\bibinfo  {journal}
  {Phys. Lett.}\ }\textbf {\bibinfo {volume} {B520}},\ \bibinfo {pages}
  {263--268} (\bibinfo {year} {2001})},\ \Eprint
  {http://arxiv.org/abs/hep-ph/0010178} {arXiv:hep-ph/0010178 [hep-ph]}
  \BibitemShut {NoStop}%
\bibitem [{\citenamefont {Strumia}(2002)}]{Strumia:2002fw}%
  \BibitemOpen
  \bibfield  {author} {\bibinfo {author} {\bibfnamefont {Alessandro}\
  \bibnamefont {Strumia}},\ }\bibfield  {title} {\enquote {\bibinfo {title}
  {{Interpreting the LSND anomaly: Sterile neutrinos or CPT violation
  or...?}}}\ }\href {\doibase 10.1016/S0370-2693(02)02042-7} {\bibfield
  {journal} {\bibinfo  {journal} {Phys. Lett.}\ }\textbf {\bibinfo {volume}
  {B539}},\ \bibinfo {pages} {91--101} (\bibinfo {year} {2002})},\ \Eprint
  {http://arxiv.org/abs/hep-ph/0201134} {arXiv:hep-ph/0201134 [hep-ph]}
  \BibitemShut {NoStop}%
\bibitem [{\citenamefont {Barenboim}\ \emph {et~al.}(2002)\citenamefont
  {Barenboim}, \citenamefont {Borissov},\ and\ \citenamefont
  {Lykken}}]{Barenboim:2002ah}%
  \BibitemOpen
  \bibfield  {author} {\bibinfo {author} {\bibfnamefont {G.}~\bibnamefont
  {Barenboim}}, \bibinfo {author} {\bibfnamefont {L.}~\bibnamefont {Borissov}},
  \ and\ \bibinfo {author} {\bibfnamefont {Joseph~D.}\ \bibnamefont {Lykken}},\
  }\bibfield  {title} {\enquote {\bibinfo {title} {{CPT violating neutrinos in
  the light of KamLAND}},}\ }\href@noop {} {\  (\bibinfo {year} {2002})},\
  \Eprint {http://arxiv.org/abs/hep-ph/0212116} {arXiv:hep-ph/0212116 [hep-ph]}
  \BibitemShut {NoStop}%
\bibitem [{\citenamefont {Gonzalez-Garcia}\ \emph {et~al.}(2003)\citenamefont
  {Gonzalez-Garcia}, \citenamefont {Maltoni},\ and\ \citenamefont
  {Schwetz}}]{GonzalezGarcia:2003jq}%
  \BibitemOpen
  \bibfield  {author} {\bibinfo {author} {\bibfnamefont {M.~C.}\ \bibnamefont
  {Gonzalez-Garcia}}, \bibinfo {author} {\bibfnamefont {M.}~\bibnamefont
  {Maltoni}}, \ and\ \bibinfo {author} {\bibfnamefont {T.}~\bibnamefont
  {Schwetz}},\ }\bibfield  {title} {\enquote {\bibinfo {title} {{Status of the
  CPT violating interpretations of the LSND signal}},}\ }\href {\doibase
  10.1103/PhysRevD.68.053007} {\bibfield  {journal} {\bibinfo  {journal} {Phys.
  Rev.}\ }\textbf {\bibinfo {volume} {D68}},\ \bibinfo {pages} {053007}
  (\bibinfo {year} {2003})},\ \Eprint {http://arxiv.org/abs/hep-ph/0306226}
  {arXiv:hep-ph/0306226 [hep-ph]} \BibitemShut {NoStop}%
\bibitem [{\citenamefont {Barger}\ \emph {et~al.}(2003)\citenamefont {Barger},
  \citenamefont {Marfatia},\ and\ \citenamefont {Whisnant}}]{Barger:2003xm}%
  \BibitemOpen
  \bibfield  {author} {\bibinfo {author} {\bibfnamefont {V.}~\bibnamefont
  {Barger}}, \bibinfo {author} {\bibfnamefont {D.}~\bibnamefont {Marfatia}}, \
  and\ \bibinfo {author} {\bibfnamefont {K.}~\bibnamefont {Whisnant}},\
  }\bibfield  {title} {\enquote {\bibinfo {title} {{LSND anomaly from CPT
  violation in four neutrino models}},}\ }\href {\doibase
  10.1016/j.physletb.2003.10.004} {\bibfield  {journal} {\bibinfo  {journal}
  {Phys. Lett.}\ }\textbf {\bibinfo {volume} {B576}},\ \bibinfo {pages}
  {303--308} (\bibinfo {year} {2003})},\ \Eprint
  {http://arxiv.org/abs/hep-ph/0308299} {arXiv:hep-ph/0308299 [hep-ph]}
  \BibitemShut {NoStop}%
\bibitem [{\citenamefont {Sorel}\ \emph {et~al.}(2004)\citenamefont {Sorel},
  \citenamefont {Conrad},\ and\ \citenamefont {Shaevitz}}]{Sorel:2003hf}%
  \BibitemOpen
  \bibfield  {author} {\bibinfo {author} {\bibfnamefont {Michel}\ \bibnamefont
  {Sorel}}, \bibinfo {author} {\bibfnamefont {Janet~M.}\ \bibnamefont
  {Conrad}}, \ and\ \bibinfo {author} {\bibfnamefont {Michael~H.}\ \bibnamefont
  {Shaevitz}},\ }\bibfield  {title} {\enquote {\bibinfo {title} {{A Combined
  analysis of short baseline neutrino experiments in the (3+1) and (3+2)
  sterile neutrino oscillation hypotheses}},}\ }\href {\doibase
  10.1103/PhysRevD.70.073004} {\bibfield  {journal} {\bibinfo  {journal} {Phys.
  Rev.}\ }\textbf {\bibinfo {volume} {D70}},\ \bibinfo {pages} {073004}
  (\bibinfo {year} {2004})},\ \Eprint {http://arxiv.org/abs/hep-ph/0305255}
  {arXiv:hep-ph/0305255 [hep-ph]} \BibitemShut {NoStop}%
\bibitem [{\citenamefont {Barenboim}\ and\ \citenamefont
  {Mavromatos}(2005)}]{Barenboim:2004wu}%
  \BibitemOpen
  \bibfield  {author} {\bibinfo {author} {\bibfnamefont {Gabriela}\
  \bibnamefont {Barenboim}}\ and\ \bibinfo {author} {\bibfnamefont {Nick~E.}\
  \bibnamefont {Mavromatos}},\ }\bibfield  {title} {\enquote {\bibinfo {title}
  {{CPT violating decoherence and LSND: A Possible window to Planck scale
  physics}},}\ }\href {\doibase 10.1088/1126-6708/2005/01/034} {\bibfield
  {journal} {\bibinfo  {journal} {JHEP}\ }\textbf {\bibinfo {volume} {01}},\
  \bibinfo {pages} {034} (\bibinfo {year} {2005})},\ \Eprint
  {http://arxiv.org/abs/hep-ph/0404014} {arXiv:hep-ph/0404014 [hep-ph]}
  \BibitemShut {NoStop}%
\bibitem [{\citenamefont {Zurek}(2004)}]{Zurek:2004vd}%
  \BibitemOpen
  \bibfield  {author} {\bibinfo {author} {\bibfnamefont {Kathryn~M.}\
  \bibnamefont {Zurek}},\ }\bibfield  {title} {\enquote {\bibinfo {title} {{New
  matter effects in neutrino oscillation experiments}},}\ }\href {\doibase
  10.1088/1126-6708/2004/10/058} {\bibfield  {journal} {\bibinfo  {journal}
  {JHEP}\ }\textbf {\bibinfo {volume} {10}},\ \bibinfo {pages} {058} (\bibinfo
  {year} {2004})},\ \Eprint {http://arxiv.org/abs/hep-ph/0405141}
  {arXiv:hep-ph/0405141 [hep-ph]} \BibitemShut {NoStop}%
\bibitem [{\citenamefont {Kaplan}\ \emph {et~al.}(2004)\citenamefont {Kaplan},
  \citenamefont {Nelson},\ and\ \citenamefont {Weiner}}]{Kaplan:2004dq}%
  \BibitemOpen
  \bibfield  {author} {\bibinfo {author} {\bibfnamefont {David~B.}\
  \bibnamefont {Kaplan}}, \bibinfo {author} {\bibfnamefont {Ann~E.}\
  \bibnamefont {Nelson}}, \ and\ \bibinfo {author} {\bibfnamefont {Neal}\
  \bibnamefont {Weiner}},\ }\bibfield  {title} {\enquote {\bibinfo {title}
  {{Neutrino oscillations as a probe of dark energy}},}\ }\href {\doibase
  10.1103/PhysRevLett.93.091801} {\bibfield  {journal} {\bibinfo  {journal}
  {Phys. Rev. Lett.}\ }\textbf {\bibinfo {volume} {93}},\ \bibinfo {pages}
  {091801} (\bibinfo {year} {2004})},\ \Eprint
  {http://arxiv.org/abs/hep-ph/0401099} {arXiv:hep-ph/0401099 [hep-ph]}
  \BibitemShut {NoStop}%
\bibitem [{\citenamefont {Pas}\ \emph {et~al.}(2005)\citenamefont {Pas},
  \citenamefont {Pakvasa},\ and\ \citenamefont {Weiler}}]{Pas:2005rb}%
  \BibitemOpen
  \bibfield  {author} {\bibinfo {author} {\bibfnamefont {Heinrich}\
  \bibnamefont {Pas}}, \bibinfo {author} {\bibfnamefont {Sandip}\ \bibnamefont
  {Pakvasa}}, \ and\ \bibinfo {author} {\bibfnamefont {Thomas~J.}\ \bibnamefont
  {Weiler}},\ }\bibfield  {title} {\enquote {\bibinfo {title} {{Sterile-active
  neutrino oscillations and shortcuts in the extra dimension}},}\ }\href
  {\doibase 10.1103/PhysRevD.72.095017} {\bibfield  {journal} {\bibinfo
  {journal} {Phys. Rev.}\ }\textbf {\bibinfo {volume} {D72}},\ \bibinfo {pages}
  {095017} (\bibinfo {year} {2005})},\ \Eprint
  {http://arxiv.org/abs/hep-ph/0504096} {arXiv:hep-ph/0504096 [hep-ph]}
  \BibitemShut {NoStop}%
\bibitem [{\citenamefont {de~Gouvêa}\ and\ \citenamefont
  {Grossman}(2006)}]{deGouvea:2006qd}%
  \BibitemOpen
  \bibfield  {author} {\bibinfo {author} {\bibfnamefont {Andre}\ \bibnamefont
  {de~Gouvêa}}\ and\ \bibinfo {author} {\bibfnamefont {Yuval}\ \bibnamefont
  {Grossman}},\ }\bibfield  {title} {\enquote {\bibinfo {title} {{A
  Three-flavor, Lorentz-violating solution to the LSND anomaly}},}\ }\href
  {\doibase 10.1103/PhysRevD.74.093008} {\bibfield  {journal} {\bibinfo
  {journal} {Phys. Rev.}\ }\textbf {\bibinfo {volume} {D74}},\ \bibinfo {pages}
  {093008} (\bibinfo {year} {2006})},\ \Eprint
  {http://arxiv.org/abs/hep-ph/0602237} {arXiv:hep-ph/0602237 [hep-ph]}
  \BibitemShut {NoStop}%
\bibitem [{\citenamefont {Schwetz}(2008)}]{Schwetz:2007cd}%
  \BibitemOpen
  \bibfield  {author} {\bibinfo {author} {\bibfnamefont {Thomas}\ \bibnamefont
  {Schwetz}},\ }\bibfield  {title} {\enquote {\bibinfo {title} {{LSND versus
  MiniBooNE: Sterile neutrinos with energy dependent masses and mixing?}}}\
  }\href {\doibase 10.1088/1126-6708/2008/02/011} {\bibfield  {journal}
  {\bibinfo  {journal} {JHEP}\ }\textbf {\bibinfo {volume} {02}},\ \bibinfo
  {pages} {011} (\bibinfo {year} {2008})},\ \Eprint
  {http://arxiv.org/abs/0710.2985} {arXiv:0710.2985 [hep-ph]} \BibitemShut
  {NoStop}%
\bibitem [{\citenamefont {Farzan}\ \emph {et~al.}(2008)\citenamefont {Farzan},
  \citenamefont {Schwetz},\ and\ \citenamefont {Smirnov}}]{Farzan:2008zv}%
  \BibitemOpen
  \bibfield  {author} {\bibinfo {author} {\bibfnamefont {Yasaman}\ \bibnamefont
  {Farzan}}, \bibinfo {author} {\bibfnamefont {Thomas}\ \bibnamefont
  {Schwetz}}, \ and\ \bibinfo {author} {\bibfnamefont {Alexei~Yu}\ \bibnamefont
  {Smirnov}},\ }\bibfield  {title} {\enquote {\bibinfo {title} {{Reconciling
  results of LSND, MiniBooNE and other experiments with soft decoherence}},}\
  }\href {\doibase 10.1088/1126-6708/2008/07/067} {\bibfield  {journal}
  {\bibinfo  {journal} {JHEP}\ }\textbf {\bibinfo {volume} {07}},\ \bibinfo
  {pages} {067} (\bibinfo {year} {2008})},\ \Eprint
  {http://arxiv.org/abs/0805.2098} {arXiv:0805.2098 [hep-ph]} \BibitemShut
  {NoStop}%
\bibitem [{\citenamefont {Hollenberg}\ \emph {et~al.}(2009)\citenamefont
  {Hollenberg}, \citenamefont {Micu}, \citenamefont {Pas},\ and\ \citenamefont
  {Weiler}}]{Hollenberg:2009ws}%
  \BibitemOpen
  \bibfield  {author} {\bibinfo {author} {\bibfnamefont {Sebastian}\
  \bibnamefont {Hollenberg}}, \bibinfo {author} {\bibfnamefont {Octavian}\
  \bibnamefont {Micu}}, \bibinfo {author} {\bibfnamefont {Heinrich}\
  \bibnamefont {Pas}}, \ and\ \bibinfo {author} {\bibfnamefont {Thomas~J.}\
  \bibnamefont {Weiler}},\ }\bibfield  {title} {\enquote {\bibinfo {title}
  {{Baseline-dependent neutrino oscillations with extra-dimensional
  shortcuts}},}\ }\href {\doibase 10.1103/PhysRevD.80.093005} {\bibfield
  {journal} {\bibinfo  {journal} {Phys. Rev.}\ }\textbf {\bibinfo {volume}
  {D80}},\ \bibinfo {pages} {093005} (\bibinfo {year} {2009})},\ \Eprint
  {http://arxiv.org/abs/0906.0150} {arXiv:0906.0150 [hep-ph]} \BibitemShut
  {NoStop}%
\bibitem [{\citenamefont {Nelson}(2011)}]{Nelson:2010hz}%
  \BibitemOpen
  \bibfield  {author} {\bibinfo {author} {\bibfnamefont {Ann~E.}\ \bibnamefont
  {Nelson}},\ }\bibfield  {title} {\enquote {\bibinfo {title} {{Effects of CP
  Violation from Neutral Heavy Fermions on Neutrino Oscillations, and the
  LSND/MiniBooNE Anomalies}},}\ }\href {\doibase 10.1103/PhysRevD.84.053001}
  {\bibfield  {journal} {\bibinfo  {journal} {Phys. Rev.}\ }\textbf {\bibinfo
  {volume} {D84}},\ \bibinfo {pages} {053001} (\bibinfo {year} {2011})},\
  \Eprint {http://arxiv.org/abs/1010.3970} {arXiv:1010.3970 [hep-ph]}
  \BibitemShut {NoStop}%
\bibitem [{\citenamefont {Akhmedov}\ and\ \citenamefont
  {Schwetz}(2010)}]{Akhmedov:2010vy}%
  \BibitemOpen
  \bibfield  {author} {\bibinfo {author} {\bibfnamefont {Evgeny}\ \bibnamefont
  {Akhmedov}}\ and\ \bibinfo {author} {\bibfnamefont {Thomas}\ \bibnamefont
  {Schwetz}},\ }\bibfield  {title} {\enquote {\bibinfo {title} {{MiniBooNE and
  LSND data: Non-standard neutrino interactions in a (3+1) scheme versus (3+2)
  oscillations}},}\ }\href {\doibase 10.1007/JHEP10(2010)115} {\bibfield
  {journal} {\bibinfo  {journal} {JHEP}\ }\textbf {\bibinfo {volume} {10}},\
  \bibinfo {pages} {115} (\bibinfo {year} {2010})},\ \Eprint
  {http://arxiv.org/abs/1007.4171} {arXiv:1007.4171 [hep-ph]} \BibitemShut
  {NoStop}%
\bibitem [{\citenamefont {Diaz}\ and\ \citenamefont
  {Kostelecky}(2011)}]{Diaz:2010ft}%
  \BibitemOpen
  \bibfield  {author} {\bibinfo {author} {\bibfnamefont {Jorge~S.}\
  \bibnamefont {Diaz}}\ and\ \bibinfo {author} {\bibfnamefont {V.~Alan}\
  \bibnamefont {Kostelecky}},\ }\bibfield  {title} {\enquote {\bibinfo {title}
  {{Three-parameter Lorentz-violating texture for neutrino mixing}},}\ }\href
  {\doibase 10.1016/j.physletb.2011.04.049} {\bibfield  {journal} {\bibinfo
  {journal} {Phys. Lett.}\ }\textbf {\bibinfo {volume} {B700}},\ \bibinfo
  {pages} {25--28} (\bibinfo {year} {2011})},\ \Eprint
  {http://arxiv.org/abs/1012.5985} {arXiv:1012.5985 [hep-ph]} \BibitemShut
  {NoStop}%
\bibitem [{\citenamefont {Bai}\ \emph {et~al.}(2016)\citenamefont {Bai},
  \citenamefont {Lu}, \citenamefont {Lu}, \citenamefont {Salvado},\ and\
  \citenamefont {Stefanek}}]{Bai:2015ztj}%
  \BibitemOpen
  \bibfield  {author} {\bibinfo {author} {\bibfnamefont {Yang}\ \bibnamefont
  {Bai}}, \bibinfo {author} {\bibfnamefont {Ran}\ \bibnamefont {Lu}}, \bibinfo
  {author} {\bibfnamefont {Sida}\ \bibnamefont {Lu}}, \bibinfo {author}
  {\bibfnamefont {Jordi}\ \bibnamefont {Salvado}}, \ and\ \bibinfo {author}
  {\bibfnamefont {Ben~A.}\ \bibnamefont {Stefanek}},\ }\bibfield  {title}
  {\enquote {\bibinfo {title} {{Three Twin Neutrinos: Evidence from LSND and
  MiniBooNE}},}\ }\href {\doibase 10.1103/PhysRevD.93.073004} {\bibfield
  {journal} {\bibinfo  {journal} {Phys. Rev.}\ }\textbf {\bibinfo {volume}
  {D93}},\ \bibinfo {pages} {073004} (\bibinfo {year} {2016})},\ \Eprint
  {http://arxiv.org/abs/1512.05357} {arXiv:1512.05357 [hep-ph]} \BibitemShut
  {NoStop}%
\bibitem [{\citenamefont {Giunti}\ and\ \citenamefont
  {Zavanin}(2015)}]{Giunti:2015mwa}%
  \BibitemOpen
  \bibfield  {author} {\bibinfo {author} {\bibfnamefont {C.}~\bibnamefont
  {Giunti}}\ and\ \bibinfo {author} {\bibfnamefont {E.~M.}\ \bibnamefont
  {Zavanin}},\ }\bibfield  {title} {\enquote {\bibinfo {title}
  {{Appearance–disappearance relation in $3 + N_s$ short-baseline neutrino
  oscillations}},}\ }\href {\doibase 10.1142/S0217732316500036} {\bibfield
  {journal} {\bibinfo  {journal} {Mod. Phys. Lett.}\ }\textbf {\bibinfo
  {volume} {A31}},\ \bibinfo {pages} {1650003} (\bibinfo {year} {2015})},\
  \Eprint {http://arxiv.org/abs/1508.03172} {arXiv:1508.03172 [hep-ph]}
  \BibitemShut {NoStop}%
\bibitem [{\citenamefont {Papoulias}\ and\ \citenamefont
  {Kosmas}(2016)}]{Papoulias:2016edm}%
  \BibitemOpen
  \bibfield  {author} {\bibinfo {author} {\bibfnamefont {D.~K.}\ \bibnamefont
  {Papoulias}}\ and\ \bibinfo {author} {\bibfnamefont {T.~S.}\ \bibnamefont
  {Kosmas}},\ }\bibfield  {title} {\enquote {\bibinfo {title} {{Impact of
  Nonstandard Interactions on Neutrino-Nucleon Scattering}},}\ }\href {\doibase
  10.1155/2016/1490860} {\bibfield  {journal} {\bibinfo  {journal} {Adv. High
  Energy Phys.}\ }\textbf {\bibinfo {volume} {2016}},\ \bibinfo {pages}
  {1490860} (\bibinfo {year} {2016})},\ \Eprint
  {http://arxiv.org/abs/1611.05069} {arXiv:1611.05069 [hep-ph]} \BibitemShut
  {NoStop}%
\bibitem [{\citenamefont {{Moss, Zander and Moulai, Marjon H. and Arg\"uelles,
  Carlos A. and Conrad, Janet M.}}(2018)}]{Moss:2017pur}%
  \BibitemOpen
  \bibfield  {author} {\bibinfo {author} {\bibnamefont {{Moss, Zander and
  Moulai, Marjon H. and Arg\"uelles, Carlos A. and Conrad, Janet M.}}},\
  }\bibfield  {title} {\enquote {\bibinfo {title} {{Exploring a nonminimal
  sterile neutrino model involving decay at IceCube}},}\ }\href {\doibase
  10.1103/PhysRevD.97.055017} {\bibfield  {journal} {\bibinfo  {journal} {Phys.
  Rev.}\ }\textbf {\bibinfo {volume} {D97}},\ \bibinfo {pages} {055017}
  (\bibinfo {year} {2018})},\ \Eprint {http://arxiv.org/abs/1711.05921}
  {arXiv:1711.05921 [hep-ph]} \BibitemShut {NoStop}%
\bibitem [{\citenamefont {Carena}\ \emph {et~al.}(2017)\citenamefont {Carena},
  \citenamefont {Li}, \citenamefont {Machado}, \citenamefont {Machado},\ and\
  \citenamefont {Wagner}}]{Carena:2017qhd}%
  \BibitemOpen
  \bibfield  {author} {\bibinfo {author} {\bibfnamefont {Marcela}\ \bibnamefont
  {Carena}}, \bibinfo {author} {\bibfnamefont {Ying-Ying}\ \bibnamefont {Li}},
  \bibinfo {author} {\bibfnamefont {Camila~S.}\ \bibnamefont {Machado}},
  \bibinfo {author} {\bibfnamefont {Pedro A.~N.}\ \bibnamefont {Machado}}, \
  and\ \bibinfo {author} {\bibfnamefont {Carlos E.~M.}\ \bibnamefont
  {Wagner}},\ }\bibfield  {title} {\enquote {\bibinfo {title} {{Neutrinos in
  Large Extra Dimensions and Short-Baseline $\nu_e$ Appearance}},}\ }\href@noop
  {} {\  (\bibinfo {year} {2017})},\ \Eprint {http://arxiv.org/abs/1708.09548}
  {arXiv:1708.09548 [hep-ph]} \BibitemShut {NoStop}%
\bibitem [{\citenamefont {Liao}\ and\ \citenamefont
  {Marfatia}(2016)}]{Liao:2016reh}%
  \BibitemOpen
  \bibfield  {author} {\bibinfo {author} {\bibfnamefont {Jiajun}\ \bibnamefont
  {Liao}}\ and\ \bibinfo {author} {\bibfnamefont {Danny}\ \bibnamefont
  {Marfatia}},\ }\bibfield  {title} {\enquote {\bibinfo {title} {{Impact of
  nonstandard interactions on sterile neutrino searches at IceCube}},}\ }\href
  {\doibase 10.1103/PhysRevLett.117.071802} {\bibfield  {journal} {\bibinfo
  {journal} {Phys. Rev. Lett.}\ }\textbf {\bibinfo {volume} {117}},\ \bibinfo
  {pages} {071802} (\bibinfo {year} {2016})},\ \Eprint
  {http://arxiv.org/abs/1602.08766} {arXiv:1602.08766 [hep-ph]} \BibitemShut
  {NoStop}%
\bibitem [{\citenamefont {Liao}\ \emph {et~al.}(2018)\citenamefont {Liao},
  \citenamefont {Marfatia},\ and\ \citenamefont {Whisnant}}]{Liao:2018mbg}%
  \BibitemOpen
  \bibfield  {author} {\bibinfo {author} {\bibfnamefont {Jiajun}\ \bibnamefont
  {Liao}}, \bibinfo {author} {\bibfnamefont {Danny}\ \bibnamefont {Marfatia}},
  \ and\ \bibinfo {author} {\bibfnamefont {Kerry}\ \bibnamefont {Whisnant}},\
  }\bibfield  {title} {\enquote {\bibinfo {title} {{MiniBooNE, MINOS+ and
  IceCube data imply a baroque neutrino sector}},}\ }\href@noop {} {\
  (\bibinfo {year} {2018})},\ \Eprint {http://arxiv.org/abs/1810.01000}
  {arXiv:1810.01000 [hep-ph]} \BibitemShut {NoStop}%
\bibitem [{\citenamefont {Asaadi}\ \emph {et~al.}(2018)\citenamefont {Asaadi},
  \citenamefont {Church}, \citenamefont {Guenette}, \citenamefont {Jones},\
  and\ \citenamefont {Szelc}}]{Asaadi:2017bhx}%
  \BibitemOpen
  \bibfield  {author} {\bibinfo {author} {\bibfnamefont {J.}~\bibnamefont
  {Asaadi}}, \bibinfo {author} {\bibfnamefont {E.}~\bibnamefont {Church}},
  \bibinfo {author} {\bibfnamefont {R.}~\bibnamefont {Guenette}}, \bibinfo
  {author} {\bibfnamefont {B.~J.~P.}\ \bibnamefont {Jones}}, \ and\ \bibinfo
  {author} {\bibfnamefont {A.~M.}\ \bibnamefont {Szelc}},\ }\bibfield  {title}
  {\enquote {\bibinfo {title} {{New light Higgs boson and short-baseline
  neutrino anomalies}},}\ }\href {\doibase 10.1103/PhysRevD.97.075021}
  {\bibfield  {journal} {\bibinfo  {journal} {Phys. Rev.}\ }\textbf {\bibinfo
  {volume} {D97}},\ \bibinfo {pages} {075021} (\bibinfo {year} {2018})},\
  \Eprint {http://arxiv.org/abs/1712.08019} {arXiv:1712.08019 [hep-ph]}
  \BibitemShut {NoStop}%
\bibitem [{\citenamefont {Döring}\ \emph {et~al.}(2018)\citenamefont
  {Döring}, \citenamefont {Päs}, \citenamefont {Sicking},\ and\ \citenamefont
  {Weiler}}]{Doring:2018cob}%
  \BibitemOpen
  \bibfield  {author} {\bibinfo {author} {\bibfnamefont {Dominik}\ \bibnamefont
  {Döring}}, \bibinfo {author} {\bibfnamefont {Heinrich}\ \bibnamefont
  {Päs}}, \bibinfo {author} {\bibfnamefont {Philipp}\ \bibnamefont {Sicking}},
  \ and\ \bibinfo {author} {\bibfnamefont {Thomas~J.}\ \bibnamefont {Weiler}},\
  }\bibfield  {title} {\enquote {\bibinfo {title} {{Sterile Neutrinos with
  Altered Dispersion Relations as an Explanation for the MiniBooNE, LSND,
  Gallium and Reactor Anomalies}},}\ }\href@noop {} {\  (\bibinfo {year}
  {2018})},\ \Eprint {http://arxiv.org/abs/1808.07460} {arXiv:1808.07460
  [hep-ph]} \BibitemShut {NoStop}%
\bibitem [{\citenamefont {Gninenko}(2009)}]{Gninenko:2009ks}%
  \BibitemOpen
  \bibfield  {author} {\bibinfo {author} {\bibfnamefont {S.~N.}\ \bibnamefont
  {Gninenko}},\ }\bibfield  {title} {\enquote {\bibinfo {title} {{The MiniBooNE
  anomaly and heavy neutrino decay}},}\ }\href {\doibase
  10.1103/PhysRevLett.103.241802} {\bibfield  {journal} {\bibinfo  {journal}
  {Phys. Rev. Lett.}\ }\textbf {\bibinfo {volume} {103}},\ \bibinfo {pages}
  {241802} (\bibinfo {year} {2009})},\ \Eprint {http://arxiv.org/abs/0902.3802}
  {arXiv:0902.3802 [hep-ph]} \BibitemShut {NoStop}%
\bibitem [{\citenamefont {Gninenko}(2011)}]{Gninenko:2010pr}%
  \BibitemOpen
  \bibfield  {author} {\bibinfo {author} {\bibfnamefont {Sergei~N.}\
  \bibnamefont {Gninenko}},\ }\bibfield  {title} {\enquote {\bibinfo {title}
  {{A resolution of puzzles from the LSND, KARMEN, and MiniBooNE
  experiments}},}\ }\href {\doibase 10.1103/PhysRevD.83.015015} {\bibfield
  {journal} {\bibinfo  {journal} {Phys. Rev.}\ }\textbf {\bibinfo {volume}
  {D83}},\ \bibinfo {pages} {015015} (\bibinfo {year} {2011})},\ \Eprint
  {http://arxiv.org/abs/1009.5536} {arXiv:1009.5536 [hep-ph]} \BibitemShut
  {NoStop}%
\bibitem [{\citenamefont {Dib}\ \emph {et~al.}(2011)\citenamefont {Dib},
  \citenamefont {Helo}, \citenamefont {Kovalenko},\ and\ \citenamefont
  {Schmidt}}]{Dib:2011jh}%
  \BibitemOpen
  \bibfield  {author} {\bibinfo {author} {\bibfnamefont {Claudio}\ \bibnamefont
  {Dib}}, \bibinfo {author} {\bibfnamefont {Juan~Carlos}\ \bibnamefont {Helo}},
  \bibinfo {author} {\bibfnamefont {Sergey}\ \bibnamefont {Kovalenko}}, \ and\
  \bibinfo {author} {\bibfnamefont {Ivan}\ \bibnamefont {Schmidt}},\ }\bibfield
   {title} {\enquote {\bibinfo {title} {{Sterile neutrino decay explanation of
  LSND and MiniBooNE anomalies}},}\ }\href {\doibase
  10.1103/PhysRevD.84.071301} {\bibfield  {journal} {\bibinfo  {journal} {Phys.
  Rev.}\ }\textbf {\bibinfo {volume} {D84}},\ \bibinfo {pages} {071301}
  (\bibinfo {year} {2011})},\ \Eprint {http://arxiv.org/abs/1105.4664}
  {arXiv:1105.4664 [hep-ph]} \BibitemShut {NoStop}%
\bibitem [{\citenamefont {McKeen}\ and\ \citenamefont
  {Pospelov}(2010)}]{McKeen:2010rx}%
  \BibitemOpen
  \bibfield  {author} {\bibinfo {author} {\bibfnamefont {David}\ \bibnamefont
  {McKeen}}\ and\ \bibinfo {author} {\bibfnamefont {Maxim}\ \bibnamefont
  {Pospelov}},\ }\bibfield  {title} {\enquote {\bibinfo {title} {{Muon Capture
  Constraints on Sterile Neutrino Properties}},}\ }\href {\doibase
  10.1103/PhysRevD.82.113018} {\bibfield  {journal} {\bibinfo  {journal} {Phys.
  Rev.}\ }\textbf {\bibinfo {volume} {D82}},\ \bibinfo {pages} {113018}
  (\bibinfo {year} {2010})},\ \Eprint {http://arxiv.org/abs/1011.3046}
  {arXiv:1011.3046 [hep-ph]} \BibitemShut {NoStop}%
\bibitem [{\citenamefont {Masip}\ \emph {et~al.}(2013)\citenamefont {Masip},
  \citenamefont {Masjuan},\ and\ \citenamefont {Meloni}}]{Masip:2012ke}%
  \BibitemOpen
  \bibfield  {author} {\bibinfo {author} {\bibfnamefont {Manuel}\ \bibnamefont
  {Masip}}, \bibinfo {author} {\bibfnamefont {Pere}\ \bibnamefont {Masjuan}}, \
  and\ \bibinfo {author} {\bibfnamefont {Davide}\ \bibnamefont {Meloni}},\
  }\bibfield  {title} {\enquote {\bibinfo {title} {{Heavy neutrino decays at
  MiniBooNE}},}\ }\href {\doibase 10.1007/JHEP01(2013)106} {\bibfield
  {journal} {\bibinfo  {journal} {JHEP}\ }\textbf {\bibinfo {volume} {01}},\
  \bibinfo {pages} {106} (\bibinfo {year} {2013})},\ \Eprint
  {http://arxiv.org/abs/1210.1519} {arXiv:1210.1519 [hep-ph]} \BibitemShut
  {NoStop}%
\bibitem [{\citenamefont {Masip}\ and\ \citenamefont
  {Masjuan}(2011)}]{Masip:2011qb}%
  \BibitemOpen
  \bibfield  {author} {\bibinfo {author} {\bibfnamefont {Manuel}\ \bibnamefont
  {Masip}}\ and\ \bibinfo {author} {\bibfnamefont {Pere}\ \bibnamefont
  {Masjuan}},\ }\bibfield  {title} {\enquote {\bibinfo {title} {{Heavy-neutrino
  decays at neutrino telescopes}},}\ }\href {\doibase
  10.1103/PhysRevD.83.091301} {\bibfield  {journal} {\bibinfo  {journal} {Phys.
  Rev.}\ }\textbf {\bibinfo {volume} {D83}},\ \bibinfo {pages} {091301}
  (\bibinfo {year} {2011})},\ \Eprint {http://arxiv.org/abs/1103.0689}
  {arXiv:1103.0689 [hep-ph]} \BibitemShut {NoStop}%
\bibitem [{\citenamefont {Gninenko}(2012)}]{Gninenko:2012rw}%
  \BibitemOpen
  \bibfield  {author} {\bibinfo {author} {\bibfnamefont {S.~N.}\ \bibnamefont
  {Gninenko}},\ }\bibfield  {title} {\enquote {\bibinfo {title} {{New limits on
  radiative sterile neutrino decays from a search for single photons in
  neutrino interactions}},}\ }\href {\doibase 10.1016/j.physletb.2012.02.071}
  {\bibfield  {journal} {\bibinfo  {journal} {Phys. Lett.}\ }\textbf {\bibinfo
  {volume} {B710}},\ \bibinfo {pages} {86--90} (\bibinfo {year} {2012})},\
  \Eprint {http://arxiv.org/abs/1201.5194} {arXiv:1201.5194 [hep-ph]}
  \BibitemShut {NoStop}%
\bibitem [{\citenamefont {Magill}\ \emph
  {et~al.}(2018{\natexlab{a}})\citenamefont {Magill}, \citenamefont {Plestid},
  \citenamefont {Pospelov},\ and\ \citenamefont {Tsai}}]{Magill:2018jla}%
  \BibitemOpen
  \bibfield  {author} {\bibinfo {author} {\bibfnamefont {Gabriel}\ \bibnamefont
  {Magill}}, \bibinfo {author} {\bibfnamefont {Ryan}\ \bibnamefont {Plestid}},
  \bibinfo {author} {\bibfnamefont {Maxim}\ \bibnamefont {Pospelov}}, \ and\
  \bibinfo {author} {\bibfnamefont {Yu-Dai}\ \bibnamefont {Tsai}},\ }\bibfield
  {title} {\enquote {\bibinfo {title} {{Dipole portal to heavy neutral
  leptons}},}\ }\href@noop {} {\  (\bibinfo {year} {2018}{\natexlab{a}})},\
  \Eprint {http://arxiv.org/abs/1803.03262} {arXiv:1803.03262 [hep-ph]}
  \BibitemShut {NoStop}%
\bibitem [{\citenamefont {Bertuzzo}\ \emph
  {et~al.}(2018{\natexlab{a}})\citenamefont {Bertuzzo}, \citenamefont {Jana},
  \citenamefont {Machado},\ and\ \citenamefont
  {Zukanovich~Funchal}}]{Bertuzzo:2018ftf}%
  \BibitemOpen
  \bibfield  {author} {\bibinfo {author} {\bibfnamefont {Enrico}\ \bibnamefont
  {Bertuzzo}}, \bibinfo {author} {\bibfnamefont {Sudip}\ \bibnamefont {Jana}},
  \bibinfo {author} {\bibfnamefont {Pedro A.~N.}\ \bibnamefont {Machado}}, \
  and\ \bibinfo {author} {\bibfnamefont {Renata}\ \bibnamefont
  {Zukanovich~Funchal}},\ }\bibfield  {title} {\enquote {\bibinfo {title}
  {{Neutrino Masses and Mixings Dynamically Generated by a Light Dark
  Sector}},}\ }\href@noop {} {\  (\bibinfo {year} {2018}{\natexlab{a}})},\
  \Eprint {http://arxiv.org/abs/1808.02500} {arXiv:1808.02500 [hep-ph]}
  \BibitemShut {NoStop}%
\bibitem [{\citenamefont {Bertuzzo}\ \emph
  {et~al.}(2018{\natexlab{b}})\citenamefont {Bertuzzo}, \citenamefont {Jana},
  \citenamefont {Machado},\ and\ \citenamefont
  {Zukanovich~Funchal}}]{Bertuzzo:2018itn}%
  \BibitemOpen
  \bibfield  {author} {\bibinfo {author} {\bibfnamefont {Enrico}\ \bibnamefont
  {Bertuzzo}}, \bibinfo {author} {\bibfnamefont {Sudip}\ \bibnamefont {Jana}},
  \bibinfo {author} {\bibfnamefont {Pedro A.~N.}\ \bibnamefont {Machado}}, \
  and\ \bibinfo {author} {\bibfnamefont {Renata}\ \bibnamefont
  {Zukanovich~Funchal}},\ }\bibfield  {title} {\enquote {\bibinfo {title} {{A
  Dark Neutrino Portal to Explain MiniBooNE}},}\ }\href@noop {} {\  (\bibinfo
  {year} {2018}{\natexlab{b}})},\ \Eprint {http://arxiv.org/abs/1807.09877}
  {arXiv:1807.09877 [hep-ph]} \BibitemShut {NoStop}%
\bibitem [{\citenamefont {Ballett}\ \emph {et~al.}(2018)\citenamefont
  {Ballett}, \citenamefont {Pascoli},\ and\ \citenamefont
  {Ross-Lonergan}}]{Ballett:2018ynz}%
  \BibitemOpen
  \bibfield  {author} {\bibinfo {author} {\bibfnamefont {Peter}\ \bibnamefont
  {Ballett}}, \bibinfo {author} {\bibfnamefont {Silvia}\ \bibnamefont
  {Pascoli}}, \ and\ \bibinfo {author} {\bibfnamefont {Mark}\ \bibnamefont
  {Ross-Lonergan}},\ }\bibfield  {title} {\enquote {\bibinfo {title} {{U(1)'
  mediated decays of heavy sterile neutrinos in MiniBooNE}},}\ }\href@noop {}
  {\  (\bibinfo {year} {2018})},\ \Eprint {http://arxiv.org/abs/1808.02915}
  {arXiv:1808.02915 [hep-ph]} \BibitemShut {NoStop}%
\bibitem [{\citenamefont {Ballett}\ \emph
  {et~al.}(2019{\natexlab{a}})\citenamefont {Ballett}, \citenamefont
  {Hostert},\ and\ \citenamefont {Pascoli}}]{Ballett:2019cqp}%
  \BibitemOpen
  \bibfield  {author} {\bibinfo {author} {\bibfnamefont {Peter}\ \bibnamefont
  {Ballett}}, \bibinfo {author} {\bibfnamefont {Matheus}\ \bibnamefont
  {Hostert}}, \ and\ \bibinfo {author} {\bibfnamefont {Silvia}\ \bibnamefont
  {Pascoli}},\ }\bibfield  {title} {\enquote {\bibinfo {title} {{Neutrino
  Masses from a Dark Neutrino Sector below the Electroweak Scale}},}\ }\href
  {\doibase 10.1103/PhysRevD.99.091701} {\bibfield  {journal} {\bibinfo
  {journal} {Phys. Rev.}\ }\textbf {\bibinfo {volume} {D99}},\ \bibinfo {pages}
  {091701} (\bibinfo {year} {2019}{\natexlab{a}})},\ \Eprint
  {http://arxiv.org/abs/1903.07590} {arXiv:1903.07590 [hep-ph]} \BibitemShut
  {NoStop}%
\bibitem [{\citenamefont {Ballett}\ \emph
  {et~al.}(2019{\natexlab{b}})\citenamefont {Ballett}, \citenamefont
  {Hostert},\ and\ \citenamefont {Pascoli}}]{Ballett:2019pyw}%
  \BibitemOpen
  \bibfield  {author} {\bibinfo {author} {\bibfnamefont {Peter}\ \bibnamefont
  {Ballett}}, \bibinfo {author} {\bibfnamefont {Matheus}\ \bibnamefont
  {Hostert}}, \ and\ \bibinfo {author} {\bibfnamefont {Silvia}\ \bibnamefont
  {Pascoli}},\ }\bibfield  {title} {\enquote {\bibinfo {title} {{Dark Neutrinos
  and a Three Portal Connection to the Standard Model}},}\ }\href@noop {} {\
  (\bibinfo {year} {2019}{\natexlab{b}})},\ \Eprint
  {http://arxiv.org/abs/1903.07589} {arXiv:1903.07589 [hep-ph]} \BibitemShut
  {NoStop}%
\bibitem [{\citenamefont {Auerbach}\ \emph {et~al.}(2001)\citenamefont
  {Auerbach} \emph {et~al.}}]{Auerbach:2001wg}%
  \BibitemOpen
  \bibfield  {author} {\bibinfo {author} {\bibfnamefont {L.~B.}\ \bibnamefont
  {Auerbach}} \emph {et~al.} (\bibinfo {collaboration} {LSND}),\ }\bibfield
  {title} {\enquote {\bibinfo {title} {{Measurement of electron - neutrino -
  electron elastic scattering}},}\ }\href {\doibase 10.1103/PhysRevD.63.112001}
  {\bibfield  {journal} {\bibinfo  {journal} {Phys. Rev.}\ }\textbf {\bibinfo
  {volume} {D63}},\ \bibinfo {pages} {112001} (\bibinfo {year} {2001})},\
  \Eprint {http://arxiv.org/abs/hep-ex/0101039} {arXiv:hep-ex/0101039 [hep-ex]}
  \BibitemShut {NoStop}%
\bibitem [{\citenamefont {Deniz}\ \emph {et~al.}(2010)\citenamefont {Deniz}
  \emph {et~al.}}]{Deniz:2009mu}%
  \BibitemOpen
  \bibfield  {author} {\bibinfo {author} {\bibfnamefont {M.}~\bibnamefont
  {Deniz}} \emph {et~al.} (\bibinfo {collaboration} {TEXONO}),\ }\bibfield
  {title} {\enquote {\bibinfo {title} {{Measurement of Nu(e)-bar -Electron
  Scattering Cross-Section with a CsI(Tl) Scintillating Crystal Array at the
  Kuo-Sheng Nuclear Power Reactor}},}\ }\href {\doibase
  10.1103/PhysRevD.81.072001} {\bibfield  {journal} {\bibinfo  {journal} {Phys.
  Rev.}\ }\textbf {\bibinfo {volume} {D81}},\ \bibinfo {pages} {072001}
  (\bibinfo {year} {2010})},\ \Eprint {http://arxiv.org/abs/0911.1597}
  {arXiv:0911.1597 [hep-ex]} \BibitemShut {NoStop}%
\bibitem [{\citenamefont {Bellini}\ \emph {et~al.}(2011)\citenamefont {Bellini}
  \emph {et~al.}}]{Bellini:2011rx}%
  \BibitemOpen
  \bibfield  {author} {\bibinfo {author} {\bibfnamefont {G.}~\bibnamefont
  {Bellini}} \emph {et~al.},\ }\bibfield  {title} {\enquote {\bibinfo {title}
  {{Precision measurement of the 7Be solar neutrino interaction rate in
  Borexino}},}\ }\href {\doibase 10.1103/PhysRevLett.107.141302} {\bibfield
  {journal} {\bibinfo  {journal} {Phys. Rev. Lett.}\ }\textbf {\bibinfo
  {volume} {107}},\ \bibinfo {pages} {141302} (\bibinfo {year} {2011})},\
  \Eprint {http://arxiv.org/abs/1104.1816} {arXiv:1104.1816 [hep-ex]}
  \BibitemShut {NoStop}%
\bibitem [{\citenamefont {Park}(2013)}]{Park:2013dax}%
  \BibitemOpen
  \bibfield  {author} {\bibinfo {author} {\bibfnamefont {Jaewon}\ \bibnamefont
  {Park}},\ }\emph {\bibinfo {title} {{Neutrino-Electron Scattering in MINERvA
  for Constraining the NuMI Neutrino Flux}}},\ \href {\doibase 10.2172/1248363}
  {Ph.D. thesis},\ \bibinfo  {school} {U. Rochester} (\bibinfo {year}
  {2013})\BibitemShut {NoStop}%
\bibitem [{\citenamefont {Valencia}\ \emph {et~al.}(2019)\citenamefont
  {Valencia} \emph {et~al.}}]{Valencia:2019mkf}%
  \BibitemOpen
  \bibfield  {author} {\bibinfo {author} {\bibfnamefont {E.}~\bibnamefont
  {Valencia}} \emph {et~al.} (\bibinfo {collaboration} {MINERvA}),\ }\bibfield
  {title} {\enquote {\bibinfo {title} {{Constraint of the MINERvA Medium Energy
  Neutrino Flux using Neutrino-Electron Elastic Scattering}},}\ }\href@noop {}
  {\  (\bibinfo {year} {2019})},\ \Eprint {http://arxiv.org/abs/1906.00111}
  {arXiv:1906.00111 [hep-ex]} \BibitemShut {NoStop}%
\bibitem [{\citenamefont {Park}\ \emph {et~al.}(2016)\citenamefont {Park} \emph
  {et~al.}}]{Park:2015eqa}%
  \BibitemOpen
  \bibfield  {author} {\bibinfo {author} {\bibfnamefont {J.}~\bibnamefont
  {Park}} \emph {et~al.} (\bibinfo {collaboration} {MINERvA}),\ }\bibfield
  {title} {\enquote {\bibinfo {title} {{Measurement of Neutrino Flux from
  Neutrino-Electron Elastic Scattering}},}\ }\href {\doibase
  10.1103/PhysRevD.93.112007} {\bibfield  {journal} {\bibinfo  {journal} {Phys.
  Rev.}\ }\textbf {\bibinfo {volume} {D93}},\ \bibinfo {pages} {112007}
  (\bibinfo {year} {2016})},\ \Eprint {http://arxiv.org/abs/1512.07699}
  {arXiv:1512.07699 [physics.ins-det]} \BibitemShut {NoStop}%
\bibitem [{\citenamefont
  {Valencia-Rodriguez}(2016)}]{Valencia-Rodriguez:2016vkf}%
  \BibitemOpen
  \bibfield  {author} {\bibinfo {author} {\bibfnamefont {Edgar}\ \bibnamefont
  {Valencia-Rodriguez}},\ }\emph {\bibinfo {title} {{Neutrino - Electron
  Scattering in MINER$\nu$A for Constraint NuMI Flux at Medium}}},\ \href
  {\doibase 10.2172/1341804} {Ph.D. thesis},\ \bibinfo  {school} {Guanajuato
  U.} (\bibinfo {year} {2016})\BibitemShut {NoStop}%
\bibitem [{\citenamefont {De~Winter}\ \emph {et~al.}(1989)\citenamefont
  {De~Winter} \emph {et~al.}}]{DeWinter:1989zg}%
  \BibitemOpen
  \bibfield  {author} {\bibinfo {author} {\bibfnamefont {K.}~\bibnamefont
  {De~Winter}} \emph {et~al.} (\bibinfo {collaboration} {CHARM-II}),\
  }\bibfield  {title} {\enquote {\bibinfo {title} {{A Detector for the Study of
  Neutrino - Electron Scattering}},}\ }\href {\doibase
  10.1016/0168-9002(89)91190-X} {\bibfield  {journal} {\bibinfo  {journal}
  {Nucl. Instrum. Meth.}\ }\textbf {\bibinfo {volume} {A278}},\ \bibinfo
  {pages} {670} (\bibinfo {year} {1989})}\BibitemShut {NoStop}%
\bibitem [{\citenamefont {Geiregat}\ \emph {et~al.}(1993)\citenamefont
  {Geiregat} \emph {et~al.}}]{Geiregat:1992zv}%
  \BibitemOpen
  \bibfield  {author} {\bibinfo {author} {\bibfnamefont {D.}~\bibnamefont
  {Geiregat}} \emph {et~al.} (\bibinfo {collaboration} {CHARM-II}),\ }\bibfield
   {title} {\enquote {\bibinfo {title} {{Calibration and performance of the
  CHARM-II detector}},}\ }\href {\doibase 10.1016/0168-9002(93)91010-K}
  {\bibfield  {journal} {\bibinfo  {journal} {Nucl. Instrum. Meth.}\ }\textbf
  {\bibinfo {volume} {A325}},\ \bibinfo {pages} {92--108} (\bibinfo {year}
  {1993})}\BibitemShut {NoStop}%
\bibitem [{\citenamefont {Vilain}\ \emph {et~al.}(1994)\citenamefont {Vilain}
  \emph {et~al.}}]{Vilain:1994qy}%
  \BibitemOpen
  \bibfield  {author} {\bibinfo {author} {\bibfnamefont {P.}~\bibnamefont
  {Vilain}} \emph {et~al.} (\bibinfo {collaboration} {CHARM-II}),\ }\bibfield
  {title} {\enquote {\bibinfo {title} {{Precision measurement of electroweak
  parameters from the scattering of muon-neutrinos on electrons}},}\ }\href
  {\doibase 10.1016/0370-2693(94)91421-4} {\bibfield  {journal} {\bibinfo
  {journal} {Phys. Lett.}\ }\textbf {\bibinfo {volume} {B335}},\ \bibinfo
  {pages} {246--252} (\bibinfo {year} {1994})}\BibitemShut {NoStop}%
\bibitem [{\citenamefont {Pospelov}\ and\ \citenamefont
  {Tsai}(2018)}]{Pospelov:2017kep}%
  \BibitemOpen
  \bibfield  {author} {\bibinfo {author} {\bibfnamefont {Maxim}\ \bibnamefont
  {Pospelov}}\ and\ \bibinfo {author} {\bibfnamefont {Yu-Dai}\ \bibnamefont
  {Tsai}},\ }\bibfield  {title} {\enquote {\bibinfo {title} {{Light scalars and
  dark photons in Borexino and LSND experiments}},}\ }\href {\doibase
  10.1016/j.physletb.2018.08.053} {\bibfield  {journal} {\bibinfo  {journal}
  {Phys. Lett.}\ }\textbf {\bibinfo {volume} {B785}},\ \bibinfo {pages}
  {288--295} (\bibinfo {year} {2018})},\ \Eprint
  {http://arxiv.org/abs/1706.00424} {arXiv:1706.00424 [hep-ph]} \BibitemShut
  {NoStop}%
\bibitem [{\citenamefont {Lindner}\ \emph {et~al.}(2018)\citenamefont
  {Lindner}, \citenamefont {Queiroz}, \citenamefont {Rodejohann},\ and\
  \citenamefont {Xu}}]{Lindner:2018kjo}%
  \BibitemOpen
  \bibfield  {author} {\bibinfo {author} {\bibfnamefont {Manfred}\ \bibnamefont
  {Lindner}}, \bibinfo {author} {\bibfnamefont {Farinaldo~S.}\ \bibnamefont
  {Queiroz}}, \bibinfo {author} {\bibfnamefont {Werner}\ \bibnamefont
  {Rodejohann}}, \ and\ \bibinfo {author} {\bibfnamefont {Xun-Jie}\
  \bibnamefont {Xu}},\ }\bibfield  {title} {\enquote {\bibinfo {title}
  {{Neutrino-electron scattering: general constraints on Z' and dark photon
  models}},}\ }\href {\doibase 10.1007/JHEP05(2018)098} {\bibfield  {journal}
  {\bibinfo  {journal} {JHEP}\ }\textbf {\bibinfo {volume} {05}},\ \bibinfo
  {pages} {098} (\bibinfo {year} {2018})},\ \Eprint
  {http://arxiv.org/abs/1803.00060} {arXiv:1803.00060 [hep-ph]} \BibitemShut
  {NoStop}%
\bibitem [{\citenamefont {Magill}\ \emph
  {et~al.}(2018{\natexlab{b}})\citenamefont {Magill}, \citenamefont {Plestid},
  \citenamefont {Pospelov},\ and\ \citenamefont {Tsai}}]{Magill:2018tbb}%
  \BibitemOpen
  \bibfield  {author} {\bibinfo {author} {\bibfnamefont {Gabriel}\ \bibnamefont
  {Magill}}, \bibinfo {author} {\bibfnamefont {Ryan}\ \bibnamefont {Plestid}},
  \bibinfo {author} {\bibfnamefont {Maxim}\ \bibnamefont {Pospelov}}, \ and\
  \bibinfo {author} {\bibfnamefont {Yu-Dai}\ \bibnamefont {Tsai}},\ }\bibfield
  {title} {\enquote {\bibinfo {title} {{Millicharged particles in neutrino
  experiments}},}\ }\href@noop {} {\  (\bibinfo {year} {2018}{\natexlab{b}})},\
  \Eprint {http://arxiv.org/abs/1806.03310} {arXiv:1806.03310 [hep-ph]}
  \BibitemShut {NoStop}%
\bibitem [{\citenamefont {Abe}\ \emph {et~al.}(2019)\citenamefont {Abe} \emph
  {et~al.}}]{Abe:2019cer}%
  \BibitemOpen
  \bibfield  {author} {\bibinfo {author} {\bibfnamefont {K.}~\bibnamefont
  {Abe}} \emph {et~al.} (\bibinfo {collaboration} {T2K}),\ }\bibfield  {title}
  {\enquote {\bibinfo {title} {{Search for neutral-current induced single
  photon production at the ND280 near detector in T2K}},}\ }\href@noop {} {\
  (\bibinfo {year} {2019})},\ \Eprint {http://arxiv.org/abs/1902.03848}
  {arXiv:1902.03848 [hep-ex]} \BibitemShut {NoStop}%
\bibitem [{\citenamefont {Formaggio}\ \emph {et~al.}(1998)\citenamefont
  {Formaggio}, \citenamefont {Conrad}, \citenamefont {Shaevitz}, \citenamefont
  {Vaitaitis},\ and\ \citenamefont {Drucker}}]{Formaggio:1998zn}%
  \BibitemOpen
  \bibfield  {author} {\bibinfo {author} {\bibfnamefont {Joseph~A.}\
  \bibnamefont {Formaggio}}, \bibinfo {author} {\bibfnamefont {Janet~M.}\
  \bibnamefont {Conrad}}, \bibinfo {author} {\bibfnamefont {Michael}\
  \bibnamefont {Shaevitz}}, \bibinfo {author} {\bibfnamefont {Artur}\
  \bibnamefont {Vaitaitis}}, \ and\ \bibinfo {author} {\bibfnamefont {Robert}\
  \bibnamefont {Drucker}},\ }\bibfield  {title} {\enquote {\bibinfo {title}
  {{Helicity effects in neutral heavy lepton decays}},}\ }\href {\doibase
  10.1103/PhysRevD.57.7037} {\bibfield  {journal} {\bibinfo  {journal} {Phys.
  Rev.}\ }\textbf {\bibinfo {volume} {D57}},\ \bibinfo {pages} {7037--7040}
  (\bibinfo {year} {1998})}\BibitemShut {NoStop}%
\bibitem [{\citenamefont {Balantekin}\ \emph {et~al.}(2018)\citenamefont
  {Balantekin}, \citenamefont {de~Gouvêa},\ and\ \citenamefont
  {Kayser}}]{Balantekin:2018ukw}%
  \BibitemOpen
  \bibfield  {author} {\bibinfo {author} {\bibfnamefont {A.~Baha}\ \bibnamefont
  {Balantekin}}, \bibinfo {author} {\bibfnamefont {André}\ \bibnamefont
  {de~Gouvêa}}, \ and\ \bibinfo {author} {\bibfnamefont {Boris}\ \bibnamefont
  {Kayser}},\ }\bibfield  {title} {\enquote {\bibinfo {title} {{Addressing the
  Majorana vs. Dirac Question with Neutrino Decays}},}\ }\href@noop {} {\
  (\bibinfo {year} {2018})},\ \Eprint {http://arxiv.org/abs/1808.10518}
  {arXiv:1808.10518 [hep-ph]} \BibitemShut {NoStop}%
\bibitem [{\citenamefont {Chun}\ \emph {et~al.}(2011)\citenamefont {Chun},
  \citenamefont {Park},\ and\ \citenamefont {Scopel}}]{Chun:2010ve}%
  \BibitemOpen
  \bibfield  {author} {\bibinfo {author} {\bibfnamefont {Eung~Jin}\
  \bibnamefont {Chun}}, \bibinfo {author} {\bibfnamefont {Jong-Chul}\
  \bibnamefont {Park}}, \ and\ \bibinfo {author} {\bibfnamefont {Stefano}\
  \bibnamefont {Scopel}},\ }\bibfield  {title} {\enquote {\bibinfo {title}
  {{Dark matter and a new gauge boson through kinetic mixing}},}\ }\href
  {\doibase 10.1007/JHEP02(2011)100} {\bibfield  {journal} {\bibinfo  {journal}
  {JHEP}\ }\textbf {\bibinfo {volume} {02}},\ \bibinfo {pages} {100} (\bibinfo
  {year} {2011})},\ \Eprint {http://arxiv.org/abs/1011.3300} {arXiv:1011.3300
  [hep-ph]} \BibitemShut {NoStop}%
\bibitem [{\citenamefont {Parke}\ and\ \citenamefont
  {Ross-Lonergan}(2015)}]{Parke:2015goa}%
  \BibitemOpen
  \bibfield  {author} {\bibinfo {author} {\bibfnamefont {Stephen}\ \bibnamefont
  {Parke}}\ and\ \bibinfo {author} {\bibfnamefont {Mark}\ \bibnamefont
  {Ross-Lonergan}},\ }\bibfield  {title} {\enquote {\bibinfo {title}
  {{Unitarity and the Three Flavour Neutrino Mixing Matrix}},}\ }\href@noop {}
  {\  (\bibinfo {year} {2015})},\ \Eprint {http://arxiv.org/abs/1508.05095}
  {arXiv:1508.05095 [hep-ph]} \BibitemShut {NoStop}%
\bibitem [{\citenamefont {Bauer}\ \emph {et~al.}(2018)\citenamefont {Bauer},
  \citenamefont {Foldenauer},\ and\ \citenamefont {Jaeckel}}]{Bauer:2018onh}%
  \BibitemOpen
  \bibfield  {author} {\bibinfo {author} {\bibfnamefont {Martin}\ \bibnamefont
  {Bauer}}, \bibinfo {author} {\bibfnamefont {Patrick}\ \bibnamefont
  {Foldenauer}}, \ and\ \bibinfo {author} {\bibfnamefont {Joerg}\ \bibnamefont
  {Jaeckel}},\ }\bibfield  {title} {\enquote {\bibinfo {title} {{Hunting All
  the Hidden Photons}},}\ }\href {\doibase 10.1007/JHEP07(2018)094} {\bibfield
  {journal} {\bibinfo  {journal} {JHEP}\ }\textbf {\bibinfo {volume} {07}},\
  \bibinfo {pages} {094} (\bibinfo {year} {2018})},\ \Eprint
  {http://arxiv.org/abs/1803.05466} {arXiv:1803.05466 [hep-ph]} \BibitemShut
  {NoStop}%
\bibitem [{\citenamefont {Jordan}\ \emph {et~al.}(2018)\citenamefont {Jordan},
  \citenamefont {Kahn}, \citenamefont {Krnjaic}, \citenamefont {Moschella},\
  and\ \citenamefont {Spitz}}]{Jordan:2018qiy}%
  \BibitemOpen
  \bibfield  {author} {\bibinfo {author} {\bibfnamefont {Johnathon~R.}\
  \bibnamefont {Jordan}}, \bibinfo {author} {\bibfnamefont {Yonatan}\
  \bibnamefont {Kahn}}, \bibinfo {author} {\bibfnamefont {Gordan}\ \bibnamefont
  {Krnjaic}}, \bibinfo {author} {\bibfnamefont {Matthew}\ \bibnamefont
  {Moschella}}, \ and\ \bibinfo {author} {\bibfnamefont {Joshua}\ \bibnamefont
  {Spitz}},\ }\bibfield  {title} {\enquote {\bibinfo {title} {{Severe
  Constraints on New Physics Explanations of the MiniBooNE Excess}},}\
  }\href@noop {} {\  (\bibinfo {year} {2018})},\ \Eprint
  {http://arxiv.org/abs/1810.07185} {arXiv:1810.07185 [hep-ph]} \BibitemShut
  {NoStop}%
\bibitem [{\citenamefont {Tanabashi}\ \emph {et~al.}(2018)\citenamefont
  {Tanabashi} \emph {et~al.}}]{Tanabashi:2018oca}%
  \BibitemOpen
  \bibfield  {author} {\bibinfo {author} {\bibfnamefont {M.}~\bibnamefont
  {Tanabashi}} \emph {et~al.} (\bibinfo {collaboration} {Particle Data
  Group}),\ }\bibfield  {title} {\enquote {\bibinfo {title} {{Review of
  Particle Physics}},}\ }\href {\doibase 10.1103/PhysRevD.98.030001} {\bibfield
   {journal} {\bibinfo  {journal} {Phys. Rev.}\ }\textbf {\bibinfo {volume}
  {D98}},\ \bibinfo {pages} {030001} (\bibinfo {year} {2018})}\BibitemShut
  {NoStop}%
\bibitem [{\citenamefont {Aliaga}\ \emph {et~al.}(2016)\citenamefont {Aliaga}
  \emph {et~al.}}]{Aliaga:2016oaz}%
  \BibitemOpen
  \bibfield  {author} {\bibinfo {author} {\bibfnamefont {L.}~\bibnamefont
  {Aliaga}} \emph {et~al.} (\bibinfo {collaboration} {MINERvA}),\ }\bibfield
  {title} {\enquote {\bibinfo {title} {{Neutrino Flux Predictions for the NuMI
  Beam}},}\ }\href {\doibase 10.1103/PhysRevD.94.092005,
  10.1103/PhysRevD.95.039903} {\bibfield  {journal} {\bibinfo  {journal} {Phys.
  Rev.}\ }\textbf {\bibinfo {volume} {D94}},\ \bibinfo {pages} {092005}
  (\bibinfo {year} {2016})},\ \bibinfo {note} {[Addendum: Phys.
  Rev.D95,no.3,039903(2017)]},\ \Eprint {http://arxiv.org/abs/1607.00704}
  {arXiv:1607.00704 [hep-ex]} \BibitemShut {NoStop}%
\bibitem [{\citenamefont {Allaby}\ \emph {et~al.}(1988)\citenamefont {Allaby}
  \emph {et~al.}}]{Allaby:1987bb}%
  \BibitemOpen
  \bibfield  {author} {\bibinfo {author} {\bibfnamefont {J.~V.}\ \bibnamefont
  {Allaby}} \emph {et~al.} (\bibinfo {collaboration} {CHARM}),\ }\bibfield
  {title} {\enquote {\bibinfo {title} {{Total Cross-sections of Charged Current
  Neutrino and Anti-neutrino Interactions on Isoscalar Nuclei}},}\ }\href
  {\doibase 10.1007/BF01584388} {\bibfield  {journal} {\bibinfo  {journal} {Z.
  Phys.}\ }\textbf {\bibinfo {volume} {C38}},\ \bibinfo {pages} {403--410}
  (\bibinfo {year} {1988})}\BibitemShut {NoStop}%
\bibitem [{\citenamefont {de~Gouvea}\ and\ \citenamefont
  {Jenkins}(2006)}]{deGouvea:2006hfo}%
  \BibitemOpen
  \bibfield  {author} {\bibinfo {author} {\bibfnamefont {Andre}\ \bibnamefont
  {de~Gouvea}}\ and\ \bibinfo {author} {\bibfnamefont {James}\ \bibnamefont
  {Jenkins}},\ }\bibfield  {title} {\enquote {\bibinfo {title} {{What can we
  learn from neutrino electron scattering?}}}\ }\href {\doibase
  10.1103/PhysRevD.74.033004} {\bibfield  {journal} {\bibinfo  {journal} {Phys.
  Rev.}\ }\textbf {\bibinfo {volume} {D74}},\ \bibinfo {pages} {033004}
  (\bibinfo {year} {2006})},\ \Eprint {http://arxiv.org/abs/hep-ph/0603036}
  {arXiv:hep-ph/0603036 [hep-ph]} \BibitemShut {NoStop}%
\bibitem [{\citenamefont {Adamson}\ \emph {et~al.}(2016)\citenamefont {Adamson}
  \emph {et~al.}}]{Adamson:2016hyz}%
  \BibitemOpen
  \bibfield  {author} {\bibinfo {author} {\bibfnamefont {P.}~\bibnamefont
  {Adamson}} \emph {et~al.} (\bibinfo {collaboration} {MINOS}),\ }\bibfield
  {title} {\enquote {\bibinfo {title} {{Measurement of single $\pi^0$
  production by coherent neutral-current $\nu$ Fe interactions in the MINOS
  Near Detector}},}\ }\href {\doibase 10.1103/PhysRevD.94.072006} {\bibfield
  {journal} {\bibinfo  {journal} {Phys. Rev.}\ }\textbf {\bibinfo {volume}
  {D94}},\ \bibinfo {pages} {072006} (\bibinfo {year} {2016})},\ \Eprint
  {http://arxiv.org/abs/1608.05702} {arXiv:1608.05702 [hep-ex]} \BibitemShut
  {NoStop}%
\bibitem [{\citenamefont {Wolcott}\ \emph {et~al.}(2016)\citenamefont {Wolcott}
  \emph {et~al.}}]{Wolcott:2016hws}%
  \BibitemOpen
  \bibfield  {author} {\bibinfo {author} {\bibfnamefont {J.}~\bibnamefont
  {Wolcott}} \emph {et~al.} (\bibinfo {collaboration} {MINERvA}),\ }\bibfield
  {title} {\enquote {\bibinfo {title} {{Evidence for Neutral-Current
  Diffractive $\pi^0$ Production from Hydrogen in Neutrino Interactions on
  Hydrocarbon}},}\ }\href {\doibase 10.1103/PhysRevLett.117.111801} {\bibfield
  {journal} {\bibinfo  {journal} {Phys. Rev. Lett.}\ }\textbf {\bibinfo
  {volume} {117}},\ \bibinfo {pages} {111801} (\bibinfo {year} {2016})},\
  \Eprint {http://arxiv.org/abs/1604.01728} {arXiv:1604.01728 [hep-ex]}
  \BibitemShut {NoStop}%
\bibitem [{\citenamefont {Olive}\ \emph {et~al.}(2014)\citenamefont {Olive}
  \emph {et~al.}}]{pdg}%
  \BibitemOpen
  \bibfield  {author} {\bibinfo {author} {\bibfnamefont {K.~A.}\ \bibnamefont
  {Olive}} \emph {et~al.} (\bibinfo {collaboration} {Particle Data Group}),\
  }\bibfield  {title} {\enquote {\bibinfo {title} {{Review of Particle
  Physics}},}\ }\href {\doibase 10.1088/1674-1137/38/9/090001} {\bibfield
  {journal} {\bibinfo  {journal} {Chin. Phys.}\ }\textbf {\bibinfo {volume}
  {C38}},\ \bibinfo {pages} {090001} (\bibinfo {year} {2014})}\BibitemShut
  {NoStop}%
\bibitem [{\citenamefont {Vilain}\ \emph {et~al.}(1999)\citenamefont {Vilain}
  \emph {et~al.}}]{Vilain:1998uw}%
  \BibitemOpen
  \bibfield  {author} {\bibinfo {author} {\bibfnamefont {P.}~\bibnamefont
  {Vilain}} \emph {et~al.} (\bibinfo {collaboration} {CHARM II}),\ }\bibfield
  {title} {\enquote {\bibinfo {title} {{Leading order QCD analysis of neutrino
  induced dimuon events}},}\ }\href {\doibase 10.1007/s100520050611} {\bibfield
   {journal} {\bibinfo  {journal} {Eur. Phys. J.}\ }\textbf {\bibinfo {volume}
  {C11}},\ \bibinfo {pages} {19--34} (\bibinfo {year} {1999})}\BibitemShut
  {NoStop}%
\bibitem [{\citenamefont {Vilain}\ \emph {et~al.}(1993)\citenamefont {Vilain}
  \emph {et~al.}}]{Vilain:1993sf}%
  \BibitemOpen
  \bibfield  {author} {\bibinfo {author} {\bibfnamefont {P.}~\bibnamefont
  {Vilain}} \emph {et~al.} (\bibinfo {collaboration} {CHARM-II}),\ }\bibfield
  {title} {\enquote {\bibinfo {title} {{Coherent single charged pion production
  by neutrinos}},}\ }\href {\doibase 10.1016/0370-2693(93)91223-A} {\bibfield
  {journal} {\bibinfo  {journal} {Phys. Lett.}\ }\textbf {\bibinfo {volume}
  {B313}},\ \bibinfo {pages} {267--275} (\bibinfo {year} {1993})}\BibitemShut
  {NoStop}%
\bibitem [{\citenamefont {Vilain}\ \emph {et~al.}(1992)\citenamefont {Vilain}
  \emph {et~al.}}]{Vilain:1992wx}%
  \BibitemOpen
  \bibfield  {author} {\bibinfo {author} {\bibfnamefont {P.}~\bibnamefont
  {Vilain}} \emph {et~al.} (\bibinfo {collaboration} {CHARM-II}),\ }\bibfield
  {title} {\enquote {\bibinfo {title} {{Neutral current coupling constants from
  neutrino and anti-neutrino - electron scattering}},}\ }\href {\doibase
  10.1016/0370-2693(92)90291-B} {\bibfield  {journal} {\bibinfo  {journal}
  {Phys. Lett.}\ }\textbf {\bibinfo {volume} {B281}},\ \bibinfo {pages}
  {159--166} (\bibinfo {year} {1992})}\BibitemShut {NoStop}%
\bibitem [{\citenamefont {Geiregat}\ \emph {et~al.}(1991)\citenamefont
  {Geiregat} \emph {et~al.}}]{Geiregat:1991md}%
  \BibitemOpen
  \bibfield  {author} {\bibinfo {author} {\bibfnamefont {D.}~\bibnamefont
  {Geiregat}} \emph {et~al.} (\bibinfo {collaboration} {CHARM-II}),\ }\bibfield
   {title} {\enquote {\bibinfo {title} {{An Improved determination of the
  electroweak mixing angle from muon-neutrino electron scattering}},}\ }\href
  {\doibase 10.1016/0370-2693(91)91665-I} {\bibfield  {journal} {\bibinfo
  {journal} {Phys. Lett.}\ }\textbf {\bibinfo {volume} {B259}},\ \bibinfo
  {pages} {499--507} (\bibinfo {year} {1991})}\BibitemShut {NoStop}%
\bibitem [{\citenamefont {Aliaga~Soplin}(2016)}]{AliagaSoplin:2016shs}%
  \BibitemOpen
  \bibfield  {author} {\bibinfo {author} {\bibfnamefont {Leonidas}\
  \bibnamefont {Aliaga~Soplin}},\ }\emph {\bibinfo {title} {{Neutrino Flux
  Prediction for the NuMI Beamline}}},\ \href {\doibase 10.2172/1250884} {Ph.D.
  thesis},\ \bibinfo  {school} {William-Mary Coll.} (\bibinfo {year}
  {2016})\BibitemShut {NoStop}%
\end{thebibliography}%

\pagebreak


\appendix

\ifx \standalonesupplemental\undefined
\setcounter{page}{1}
\setcounter{figure}{0}
\setcounter{table}{0}
\setcounter{equation}{0}
\fi

\renewcommand{\thepage}{Supplementary Methods and Tables -- S\arabic{page}}
\renewcommand{\figurename}{SUPPL. FIG.}
\renewcommand{\tablename}{SUPPL. TABLE}

\renewcommand{\theequation}{A\arabic{equation}}
\clearpage

\begin{center}
\textbf{\large Supplementary Material}
\end{center}

Our analysis discussed in the main text is now described in more detail and all assumptions used in our simulations are summarized.
We start by discussing our statistical method, and then discuss our Monte Carlo (MC) simulation, stating our signal definitions more precisely.
Later, we show a few kinematical distributions from our dedicated MC simulation, including the angular distributions at MiniBooNE used to obtain the vertical lines in~\reffig{fig:final_plot}.
In order to furher aid reproducibility of our result, we also make our Monte Carlo events for some parameter choices available on \textsc{g}it\textsc{h}ub~\footnote{\href{https://github.com/mhostert/DarkNews}{\color{BlueViolet}\faGithub\,\,{github.com/mhostert/DarkNews}}.}.

\section{Statistical Analysis}

Our statistical analysis uses the Pearson-$\chi^2$ as a test statistic, where the expected number of events is scaled by nuisance parameters to incorporate systematic uncertainties.
Our test statistic reads
\begin{widetext}
\begin{eqnarray}
\chi^2(\vec\theta, \alpha, \beta) = \frac{ (N_{\rm data} - N_\mathrm{pred}(\vec\theta, \alpha, \beta) )^2 }{ N_\mathrm{pred}(\vec\theta, \alpha, \beta)} + \left(\frac{\alpha}{\sigma_\alpha}\right)^2  +
\left(\frac{\beta}{\sigma_\beta}\right)^2,
\end{eqnarray}
with the number of predicted events given by \begin{eqnarray}
N_\mathrm{pred}(\vec\theta, \alpha, \beta) = (1+\alpha + \beta) \mu_{\rm MC}^{\rm BKG} + (1+\alpha) \mu_{\rm MC}^{\nu-e} + (1 + \alpha) \mu_{\rm BSM}(\vec\theta),
\end{eqnarray}
\end{widetext}
where $\vec\theta$ are the model parameters, while $\alpha$ and $\beta$ are nuisance parameters that incorporate uncertainties from the overall rate and the background rate, respectively.
Here, $N_{\rm data}$ stands for the total rate observed in the experiment, $\mu_{\rm MC}^{\rm BKG}$ the quoted background rates, $\mu_{\rm MC}^{\nu-e}$ the quoted $\nu-e$ events, and $\mu_{\rm BSM}(\vec\theta)$ the predicted number of BSM events calculated using our MC.
We discuss the choice of these systematic uncertainties, namely the choice of $\sigma_\alpha$ and $\sigma_\beta$, when describing the simulation of each experiment below.
To obtain our results we use the test statistic profiled over the nuisance parameters, namely $\chi^2(\vec\theta) = \min_{(\alpha, \beta)} \left(\chi^2(\vec\theta, \alpha, \beta) \right)$, and use the test-statistic thresholds given in~\cite{pdg}.

\section{Simulation Details}
\renewcommand{\arraystretch}{1.0}
\begin{table*}[t]
    \centering
    \begin{tabular}{lp{5.2cm}p{2.8cm}p{4cm}}
        \hline        \hline
        Experiment & Detector Resolution& Overlapping & Analysis Cuts  \\
        \hline        \hline
        MiniBooNE \\        \hline
         & $\sigma_E/E = 12\%/\sqrt{E_e/{\rm GeV}}$ \newline $\sigma_\theta = 4{^\circ}$ & $E_{+} > 30$ MeV \newline $E_{-} > 30$ MeV \newline $\Delta \theta_{\pm} < 13^\circ$ &        N/A
        \\        \hline

        MINER$\nu$A
        \\        \hline
        &  $\sigma_E/E = 5.9\%/\sqrt{E_e/{\rm GeV}} + 3.4\%$ \newline $\sigma_\theta = 1{^\circ}$ & $E_{+} > 30$ MeV \newline $E_{-} > 30$ MeV \newline $\Delta \theta_{\pm} < 8^\circ$  & $E_{\rm vis} > 0.8$ GeV \newline $E_{\rm vis} \theta^2 < 3.2$ MeV
        \newline $Q^2_{\rm rec} < 0.02$ GeV$^2$\newline 
        \\        \hline

        CHARM-II & \\        \hline
         & $\sigma_E/E = 9\%/\sqrt{E/{\rm GeV}} +  11\%$ \newline $\sigma_\theta/{\rm mrad} =  \frac{27 (E/{\rm GeV})^2 +14}{\sqrt{E/{\rm GeV}}} + 1$ & $E_{+} > 30$ MeV \newline $E_{-} > 30$ MeV \newline $\Delta \theta_{\pm} < 4^\circ$ & $3$ GeV $<E_{\rm vis} < 24$ GeV \newline $E_{\rm vis} \theta^2 < 28$ MeV
        \newline   \\
    \hline        \hline
    \end{tabular}
    \caption{Experimental resolution, condition for dielectrons to be reconstructed as overlapping EM showers, and analysis cuts for the detectors studied in the main text.}
    \label{tab:parameters}
\end{table*}

We generate events distributed according to the upscattering cross section for the process $\nu_\mu A \to \nu_4 A$, where $A$ is a nuclear target.
We only discuss upscattering on nuclei, as the number of incoherent scattering on protons is much smaller for the relevant $Z^\prime$ masses; see \reffig{fig:cross_section}.
We then implement the chain of two-body decays: $\nu_4 \to \nu_\mu Z^\prime$ followed by $Z^\prime \to e^+ e^-$.
To go from our MC true quantities to the predicted experimental observables, we perform three procedures.
First, we smear the energy and angles of the $e^+$ and $e^-$ originating from the decay of the $Z^\prime$ according to detector-dependent Gaussian resolutions.
Next, we select all events with that satisfy the $e^+e^-$ overlapping condition given in~\reftab{tab:parameters}.
Namely, if the condition is satisfied they are assumed to be reconstructed as a single electromagnetic (EM) shower.
This guarantees that the events behave like a photon shower inside the detector~\footnote{For MiniBooNE, we also include events that are highly asymmetric in energy, \textit{i.e.}, $E_{\pm} > 30$ MeV and $E_{\mp} < 30$ MeV, where the most energetic shower defines the angle with respect to the beam.}.
Finally, for \minerva and CHARM-II, these samples are subject to analysis-dependent kinematic cuts to determine if they contribute to the $\nu-e$ scattering sample.
Detector resolutions, requirements for the dielectron pair to be overlapping, and analysis-dependent cuts are summarized in \reftab{tab:parameters}.
We now list the experimental parameters used in our simulations for each individual detector.

\paragraph{CHARM-II} The CHARM-II experiment is simulated using the CERN West Area Neutrino Facility (WANF) wide band beam ~\cite{Vilain:1998uw}.
The total number of protons-on-target (POT) is $2.5 \times 10^{19}$ for the $\nu$ and $\overline{\nu}$ runs combined.
We assume glass to be the main detector material, (SiO$_2$), such that we can treat neutrino scattering off an average target with $\langle Z\rangle=11$ and $\langle A \rangle = 20.7$~\cite{DeWinter:1989zg,Vilain:1993sf}.
The fiducial volume in our analysis is confined to a transverse area of $320$cm$^2$, which corresponds to a fiducial mass of $547$t, and the detection efficiency is taken to be $76\%$; efficiency for $\pi^0$ sample is quoted at $82\%$~\cite{Vilain:1992wx}.
We reproduce the total number of $\nu-e$ scattering events with $3$ GeV $< E_{\rm vis} <24$ GeV, namely $2677+2752$, to within a few percent level when setting the number of POTs in $\nu$ mode to be $1.69$ of that in the $\overline{\nu}$ mode~\cite{Geiregat:1991md}.
We assume a flux uncertainty of $\sigma_\alpha = 4.7\%$ for neutrino, and $\sigma_\alpha = 5.2\%$ for antineutrino beam~\cite{Vilain:1992wx}.
The background uncertainty is constrained to be $\sigma_\beta = 3\%$ using the data with $E_{\rm vis} \theta^2 > 28$ MeV, where the number of new physics events is negligible.

\paragraph{MINER$\nu$A} For our MINER$\nu$A simulation, we use the low-energy (LE) and medium-energy (ME) NuMI neutrino fluxes~\cite{AliagaSoplin:2016shs}.
The total number of POT is $3.43\times 10^{20}$ for LE data, and $11.6\times10^{20}$ for ME data. The detector is assumed to be made of CH, with a fiducial mass of $6.10$ tons and detection efficiencies of $73\%$~\cite{Parke:2015goa,Valencia:2019mkf}.
We assume a flux uncertainty of $\sigma_\alpha = 10\%$ for both the LE and ME modes~\cite{Aliaga:2016oaz}.
Due to the tuning performed in the sideband of interest, the uncertainties on the background rate are much larger.
For the LE, we take $\sigma_\beta = 30\%$, while for the ME data  $\sigma_\beta = 50\%$.
Although tuning is significant for the coherent $\pi^0$ production sample, the overall rate of backgrounds in the sideband with large $dE/dx$ does not vary by more than $20\%$ ($40\%$) in the LE (ME) tuning.  

\paragraph{MiniBooNE} To simulate MiniBooNE, we use the Booster Neutrino Beam (BNB) fluxes from~\cite{AguilarArevalo:2008yp}.
Here, we only discuss the neutrino run, although the predictions for the antineutrino run are very similar.
We assume a total of $12.84 \times 10^{20}$ POT in neutrino mode.
The fiducial mass of the detector is taken as $450$t of CH$_2$.
In order to apply detector efficiencies, we compute the reconstructed neutrino energy under the assumption of CCQE scattering
\begin{align}
 E_\nu^{CCQE} = \frac{E_{\rm vis} m_p}{m_p - E_{\rm vis} (1 - \cos{\theta}) },
\end{align}
where $E_{\rm vis} = E_{e^+} + E_{e^-}$ is the total visible energy after smearing.
Under this assumption, we can apply the efficiencies provided by the MiniBooNE collaboration~\cite{Aguilar-Arevalo:2012fmn}.
Using our MC we can reproduce well the distributions obtained using the MiniBooNE Monte Carlo data release provided for oscillation analyses.

\section{Kinematic Distributions}
\begin{figure*}[h]
    \centering
    \includegraphics[width=0.49\textwidth]{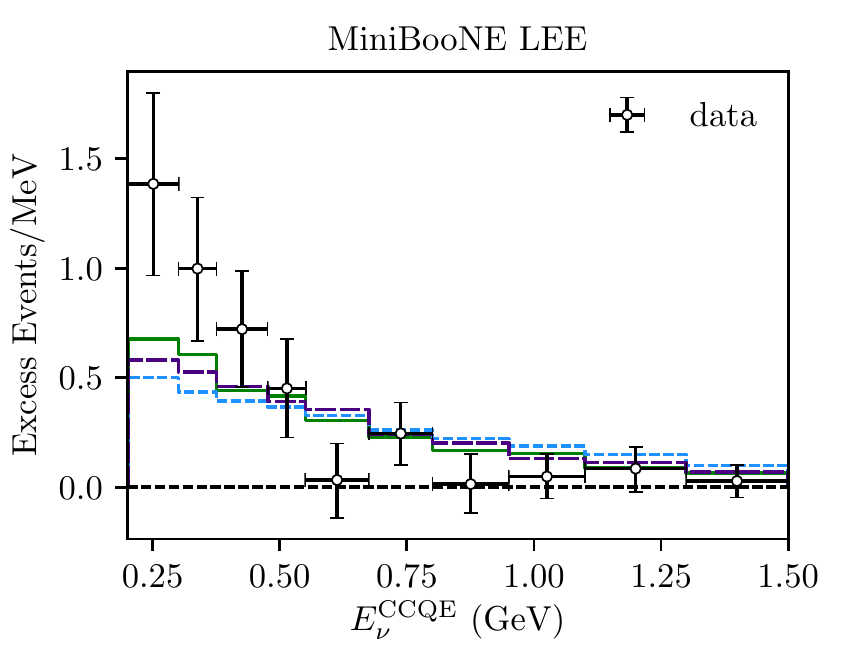}
    \includegraphics[width=0.49\textwidth]{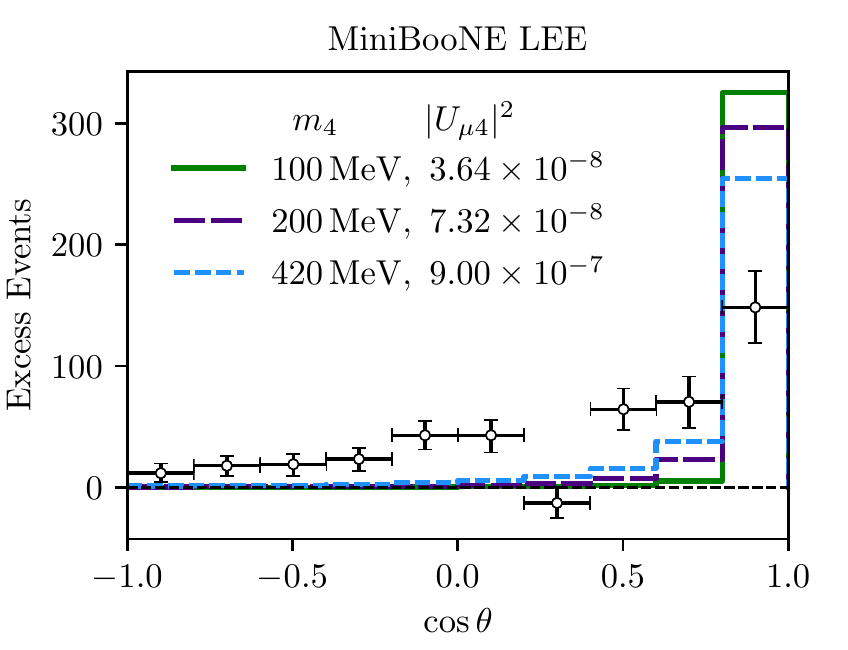}
    \caption{Data and prediction for the reconstructed neutrino energy at MiniBooNE under the assumption of CCQE scattering (\textbf{left}), and for the cosine of the angle between the visible EM signal and the neutrino beam (\textbf{right}).~\label{fig:MB_distributions}}
\end{figure*}

As an important check of our calculation and of the explanation of the MiniBooNE excess within the model of interest, we plot the MiniBooNE neutrino data from 2018~\cite{Aguilar-Arevalo:2018gpe} against our MC prediction in Suppl. Fig.~\ref{fig:MB_distributions}.
We do this for three different new physics parameter choices.
We set $m_{Z^\prime} = 30$ MeV, $\alpha \epsilon^2 = 2\times10^{-10}$ and $\alpha_D = 1/4$ for all points, but vary $|U_{\mu 4}|^2$ and $m_4$ so that the final number of excess events predicted by the model at MiniBooNE equals 334.

\begin{figure*}[h]
    \centering
    \includegraphics[width=0.49\textwidth]{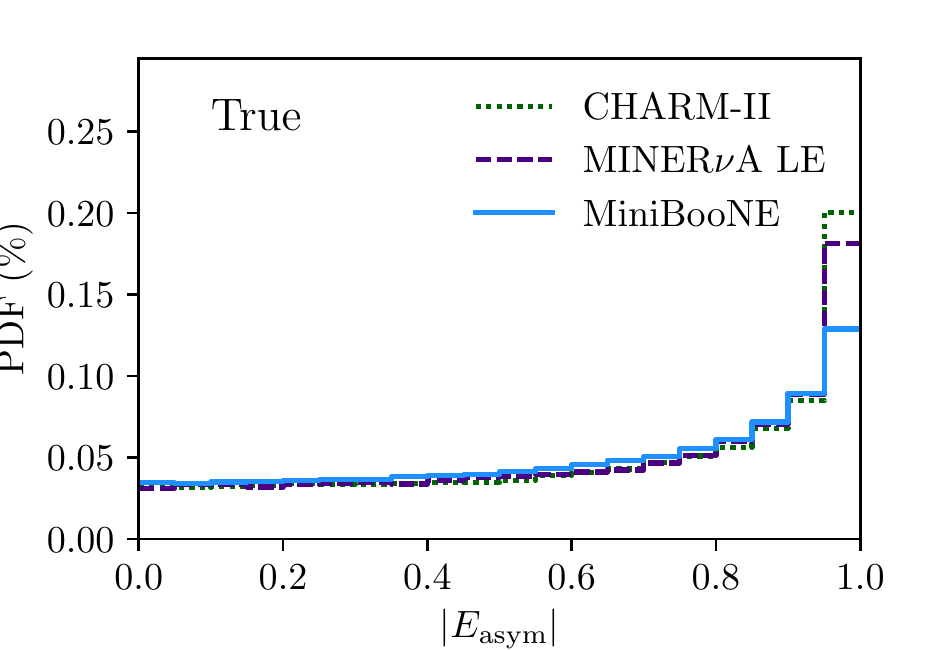}
    \includegraphics[width=0.49\textwidth]{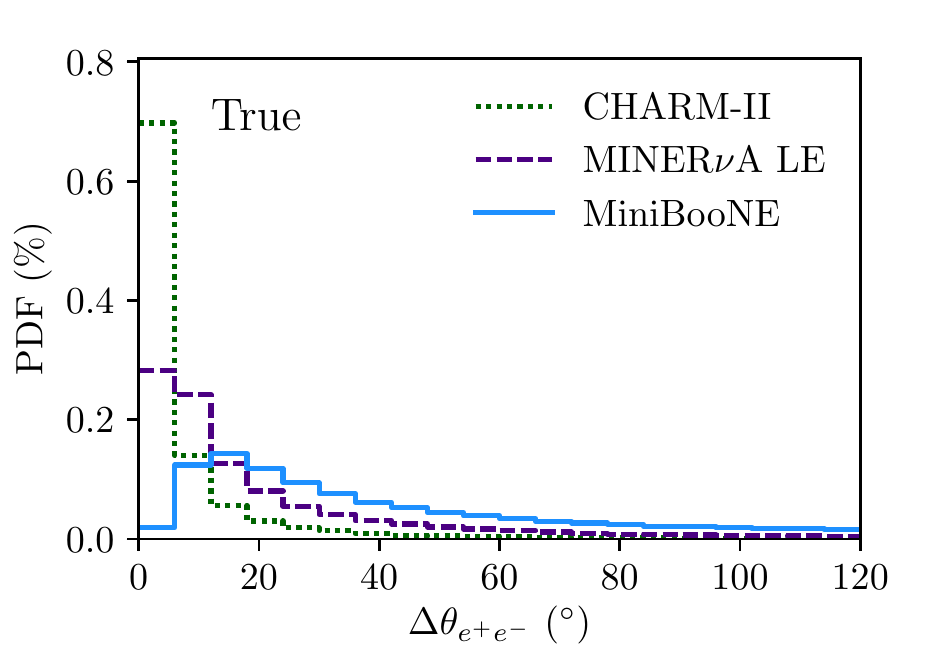}
    \\
    \includegraphics[width=0.49\textwidth]{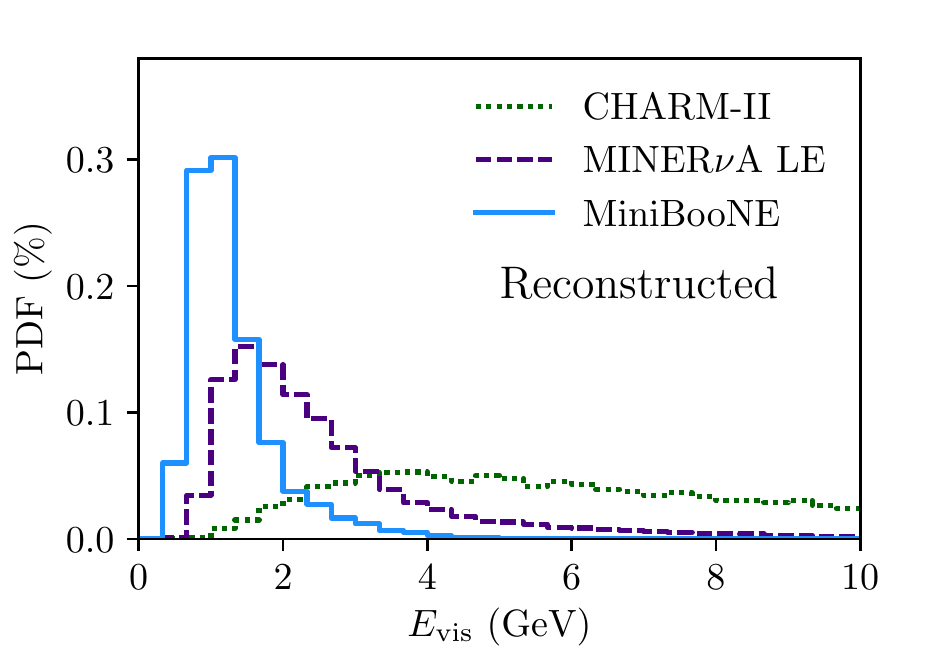}
    \includegraphics[width=0.49\textwidth]{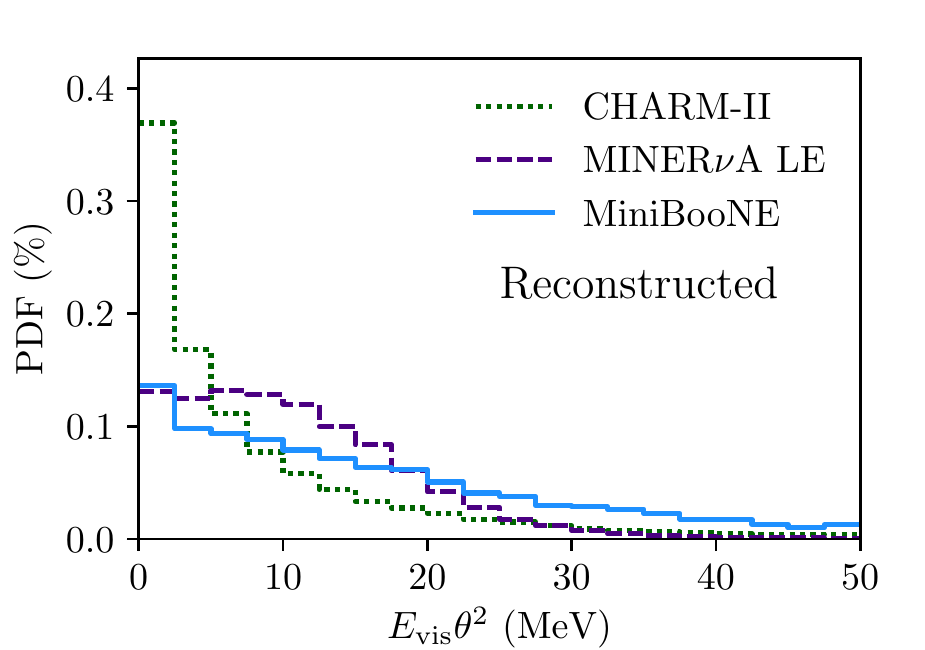}
    \caption{Kinematical distributions for the new physics events at CHARM-II, MINER$\nu$A LE and MiniBooNE for the BP.
    We show the energy asymmetry (\textbf{top left}), the electron separation angles (\textbf{top right}), both at MC truth level.
    We also show reconstructed ({after smearing}) total visible energy $E_{\rm vis}$ (\textbf{bottom left}) and $E_{\rm vis} \theta^2$ (\textbf{bottom right}).\label{fig:other_distributions}}
\end{figure*}

To verify that the new physics signal is important in neutrino-electron studies, we also plot kinematical distributions for the benchmark point (BP) discussed in the main text for different detectors.
This corresponds to $m_{Z^\prime} = 30$ MeV, $\alpha \epsilon^2 = 2\times10^{-10}$, $\alpha_D = 1/4$, $|U_{\mu 4}|^2 = 9\times10^{-7}$ and $m_4 = 420$ MeV.
The interesting variables are the energy asymmetry of the dielectron pair 
\begin{equation}
    |E_{\rm asym}| = \frac{|E_+ - E_-|}{E_+ + E_-},
\end{equation}
as well as the separation angle $\Delta \theta_{e^+e^-}$ between the two electrons.
These variables are plotted in Suppl. Fig.~\ref{fig:other_distributions} at MC truth level, before any smearing or selection takes place.
We also plot the total reconstructed energy $E_{\rm vis} = E_{e^+} + E_{e^-}$ and the quantity $E_{\rm vis} \theta^2$, where $\theta$ stands for the angle formed by the reconstructed EM shower and the neutrino beam.
The visible energy, $E_{\rm vis}$, and angle, $\theta$, are computed after smearing, but before the selection into overlapping pairs takes place.
\end{document}